\documentclass{aa}
\usepackage{graphicx}
\usepackage{natbib}
\bibpunct{(}{)}{;}{a}{}{,}
\begin{document}
%
\newcommand{\be}{\begin{equation}}
\newcommand{\ee}{\end{equation}}
\newcommand{\bea}{\begin{eqnarray}}
\newcommand{\eea}{\end{eqnarray}}
\newcommand{\Sun}{_{\sun}}
\newcommand{\Rot}{_{\mathrm{rot}}}
\newcommand{\Therm}{_{\mathrm{therm}}}
\newcommand{\Max}{_{\mathrm{max}}}
\newcommand{\Min}{_{\mathrm{min}}}
\newcommand{\Pot}{_{\mathrm{pot}}}
\newcommand{\Kin}{_{\mathrm{kin}}}
\newcommand{\Tot}{_{\mathrm{tot}}}
\newcommand{\Gal}{_{\mathrm{gal}}}
\newcommand{\ICM}{_{\mathrm{ICM}}}
\newcommand{\ISM}{_{\mathrm{ISM}}}
\newcommand{\DM}{_{\mathrm{DM}}}
\newcommand{\DMnull}{_{\mathrm{0,DM}}}
\newcommand{\Gas}{_{\mathrm{gas}}}
\newcommand{\Stars}{_*}
\newcommand{\Bulge}{_{\mathrm{bulge}}}
\newcommand{\Starsnull}{_{\mathrm{0,stars}}}
\newcommand{\Ram}{_{\mathrm{ram}}}
\newcommand{\Switch}{_{\mathrm{switch}}}
\newcommand{\Strip}{_{\mathrm{strip}}}
\newcommand{\Fallback}{_{\mathrm{fb}}}
\newcommand{\Stay}{_{\mathrm{stay}}}
\newcommand{\Origreg}{_{\mathrm{orig}}}
\newcommand{\Cylreg}{_{\mathrm{cyl}}}
\newcommand{\Bound}{_{\mathrm{bnd}}}
\newcommand{\Visc}{_{\mathrm{visc}}}
\newcommand{\Grav}{_{\mathrm{grav}}}
\newcommand{\Disk}{_{\mathrm{disk}}}

\newcommand{\const}{\textrm{const.}}
\newcommand{\degree}{^o}
\newcommand{\K}{\,\textrm{K}}
\newcommand{\Kpc}{\,\textrm{kpc}}
\newcommand{\Mpc}{\,\textrm{Mpc}}
\newcommand{\PC}{\,\textrm{pc}}
\newcommand{\Yr}{\,\textrm{yr}}
\newcommand{\CM}{\,\textrm{cm}}
\newcommand{\Sec}{\,\textrm{s}}
\newcommand{\Myr}{\,\textrm{Myr}}
\newcommand{\Gyr}{\,\textrm{Gyr}}
\newcommand{\Kms}{\,\textrm{km}\,\textrm{s}^{-1}}
\newcommand{\Cm}{\,\textrm{cm}}
\newcommand{\Erg}{\,\textrm{erg}}
\newcommand{\ccm}{\,\textrm{cm}^{-3}}
\newcommand{\gccm}{\,\textrm{g}\,\textrm{cm}^{-3}}
\newcommand{\Presunit}{\,\textrm{erg}\,\textrm{cm}^{-3}}
\newcommand{\Rampresunit}{\,\textrm{cm}^{-3}\,\textrm{km}^2\,\textrm{s}^{-2}}

%
\title{Ram pressure stripping of disk galaxies}
\subtitle{From high to low density environments}

\author{Elke Roediger
\inst{1}
\and
Gerhard Hensler\inst{2}
}

\offprints{Elke Roediger}

\institute{Institute of Theoretical Physics and Astrophysics,
           University of Kiel,
           Olshausenstr. 40, D-24098 Kiel, Germany\\
    \email{elke@astrophysik.uni-kiel.de}
    \and
          Institute of Astronomy, University of Vienna,
          T\"urkenschanzstrasse 17, A-1180 Vienna, Austria\\
   \email{hensler@astro.univie.ac.at}
}

\date{Received; accepted}

\abstract{
Galaxies in clusters and groups moving through the intracluster or
intragroup medium (abbreviated ICM for both) are expected to lose at
least a part of their interstellar medium (ISM) by the ram pressure
they experience.  We perform high resolution 2D hydrodynamical
simulations of face-on ram pressure stripping (RPS) of disk galaxies
to compile a comprehensive parameter study varying galaxy properties
(mass, vertical structure of the gas disk) and covering a large range
of ICM conditions, reaching from high density environments like in
cluster centres to low density environments typical for cluster
outskirts or groups. We find that the ICM-ISM interaction proceeds in
three phases: firstly the instantaneous stripping phase, secondly the
dynamic intermediate phase, thirdly the quasi-stable continuous
viscous stripping phase. In the first phase (time scale 20 to
$200\Myr$) the outer part of the gas disk is displaced but only
partially unbound. In the second phase (10 times as long as the first
phase) a part of the displaced gas falls back (about 10\% of the
initial gas mass) despite the constant ICM wind, but most displaced
gas is now unbound. In the third phase the galaxy continues to lose
gas at a rate of about $1\,M\Sun\,\Yr^{-1}$ by turbulent viscous
stripping. We find that the stripping efficiency depends slightly on
the Mach number of the flow, however, the main parameter is the ram
pressure. The stripping efficiency does not depend on the vertical
structure and thickness of the gas disk.

We discuss uncertainties in the classic estimate of the stripping
radius of \citet{gunn72}, which compares the ram pressure to the
gravitational restoring force. In addition, we adapt the estimate used
by \cite{mori00} for spherical galaxies, namely the comparison of the
central pressure with ram pressure. We find that the latter estimate
predicts the radius and mass of the gas disk remaining at the end of
the second phase very well, and better than the \citet{gunn72}
criterion.

From our simulations we conclude that gas disks of galaxies in high
density environments are heavily truncated or even completely
stripped, but also the gas disks of galaxies in low density
environments are disturbed by the flow and back-falling material, so
that they should also be pre-processed.
}
\maketitle
%
%
%
%
%
%
\section{Introduction}
%
The differences observed between disk galaxies in clusters and in the
field indicate that the evolution of galaxies is not only ruled by
interior processes, but also strongly by the galaxy's
environment. Some cluster galaxies show distortions in their gas
components as well in as their stellar disks. These examples can be
explained by tidal interactions with some fellow cluster members
(mergers or harassment, see review by \citealt{mihos04} and references
therein). However, a number of galaxies have rather normal stellar
disks but truncated gas disks (e.g.~NGC 4522:
\citealt{kenney99,kenney01},\citealt{vollmer04a}; NGC 4548:
\citealt{vollmer99}; NGC 4848:
\citealt{vollmer01}). \citet{cayatte90,cayatte94} observed the
brightest spiral galaxies in the Virgo cluster and found the trend
that the galaxies closest to the cluster centre have the smallest HI
disks. In general, cluster spirals are HI deficient compared to their
field relatives. They are more deficient near the cluster centre
\citep{solanes01}, but deficient galaxies can be observed out to large
distances (2 Abell radii) from the cluster centre
\citep{solanes01}. HI deficient galaxies tend to be on more radial
orbits which lead them deeper into the
cluster. \citet{koopmann98,koopmann04b,koopmann04a} studied the star
formation rates in Virgo spirals and found that also the star
formation of many cluster members is truncated, that means it is
suppressed significantly in the outer regions, but normal or even
enhanced in the central region of the galaxy.

The combination of an intact stellar disk but truncated gas disk can
hardly be achieved by tidal interactions, as this usually also
disturbs the stellar component. But even if a cluster galaxy does not
interact with another galaxy, it still moves through the intra-cluster
medium (ICM) and experiences a ram pressure $p\Ram\approx\rho\ICM\cdot
v\ICM^2$, that depends on the ICM density $\rho\ICM$ and the relative
velocity $v\ICM$ between the ICM and the galaxy. What is expected to
happen to such a galaxy? For a face-on motion, \citet{gunn72}
suggested to compare $p\Ram$ and the gravitational restoring force per
unit area $\frac{\partial\Phi}{\partial z}\Sigma\Gas(r)$, which
depends on the gradient of the galactic potential $\Phi$ in the
direction perpendicular to the disk ($z$-direction) and the surface
density of the gas disk $\Sigma\Gas$. At radii $r$ where
\be
p\Ram > \frac{\partial\Phi}{\partial z}(r)\,\Sigma\Gas(r), \label{eq:gunngott}
\ee
the gas will be pushed out of the disk. Applying this estimate to
typical disk galaxies in clusters, galaxies passing near the cluster
core should get stripped of large parts of their gas disks. In
addition to the ram pressure stripping (RPS), a remaining inner gas
disk should be compressed.

As the common interest concentrates on the effect on the galaxy,
numerical simulations of RPS are usually done in the rest frame of the
galaxy. Here the relative motion translates into an ICM wind.

First attempts to simulate RPS of disk galaxies have been made by
\citet{farouki80} and \citet{toyama80}, but these works were limited
to low resolution and short run times due to weaker computational
power. Later studies with better resolution (e.g. SPH simulations by
\citealt{abadi99}, \citealt{schulz01} and 3D Eulerian hydro
simulations by \citealt{quilis00}) performed a few simulations of
massive disk galaxies in constant mostly face-on winds. The simulated
winds were representative for the conditions in cluster centres, the
galaxies suffered substantial gas loss. In general, previous works
found that the gas disks are stripped on a short time scale (few
$10\Myr$ to few $100\Myr$), and that the radius of the remaining gas
disk is in good agreement with the estimate derived from
Eq.~\ref{eq:gunngott}. Both \citet{quilis00} and \citet{schulz01} used
longer run times and found a long phase of turbulent viscous stripping
following the rather short RPS phase. This viscous stripping works on
galaxies that are oriented inclined or edge-on towards the ICM wind as
well as on face-on oriented ones. The RPS phase seems to take longer
for inclined cases, but it can be suppressed completely for strict
edge-on cases only
\citep{quilis00,schulz01,marcolini03}. \citet{quilis00} emphasised
that the gas disks are not homogeneous but porous and tried to model
this feature by holes in the disk. In such cases the turbulent viscous
stripping was more effective. \citet{vollmer01a} applied a sticky
particle code to simulate the passage of disk galaxies through
clusters, stressing the fact that in reality the galaxies experience a
rapidly changing ICM wind instead of a constant one. Consequently, the
stripping is a distinct event. Moreover, during this event not all
displaced gas is unbound, and up to 10\% of the initial gas mass fall
back into the disk. In further papers
\citep{vollmer00,vollmer01,vollmer99,vollmer03} this group tried to
explain the features of individual galaxies by comparing simulation
results with observations and specifying the ram pressure, inclination
angle and the moment of the stripping event. \citet{marcolini03}
simulated RPS of disky dwarf galaxies in poor groups, where the ICM
densities and hence ram pressures are weaker than in clusters.

Most previous works performed a few simulation runs with massive
galaxies in winds representative for cluster centres and found that
for these cases the gas loss can be predicted by Eq.~\ref{eq:gunngott}
(at least for face-on winds). Massive galaxies in weak winds
corresponding to conditions in cluster outskirts were not simulated,
as in general it is assumed that weak winds do not have important
effects. Here we present a comprehensive parameter study of massive to
medium-mass disk galaxies in face-on ICM winds which are
representative for the whole range from cluster centres to cluster
outskirts and groups. We want to check if we can confirm the
analytical estimate for this large range of ram pressures. Moreover,
the analytical estimate incorporates only $p\Ram$ and does not
distinguish with which combination of $\rho\ICM$ and $v\ICM$ this
$p\Ram$ was achieved. We want to test systematically if the analytical
estimate misses some dependence on Mach number, especially as $v\ICM$
is expected to be in the transonic range. As a second point, we want
to investigate the dependence of RPS on galactic
parameters. Concerning the dependence on overall galaxy mass, we
perform simulations for a massive and a medium size disk galaxy. In
addition, in this paper we pay special attention to the vertical
structure and thickness of the gas disk, as the analytical estimate
(Eq.~\ref{eq:gunngott}) does not distinguish such cases, either. The
influence of an inhomogeneous gas disk will be studied elsewhere
(Roediger \& Hensler, in preparation).

The outline of this paper is as follows: In Sect.~\ref{sec:numerics}
we briefly introduce our code. In Sect.~\ref{sec:initial} we describe
our initial model. Section~\ref{sec:simulations} gives an overview of
the simulations we performed and explains some analysis
techniques. Section~\ref{sec:results} presents the results and in
Sect.~\ref{sec:discussion} we discuss them and draw conclusions.

Throughout the paper we will use $r$ and $z$ for the radial and axial
cylindrical coordinates and $R$ for the spherical radius.
%
%

\section{Numerics}
\label{sec:numerics}
To answer our questions we perform simulations using a 2D Eulerian
hydrodynamics code (assuming cylindrical symmetry) developed from an
original implementation of \citet{rozyczka85}. The restriction to 2D
reduces the computational effort significantly, allowing to
perform a large parameter study. The code is very similar to the one
described in \citet{norman86} and the ZEUS code \citep{stone92}.
Detailed descriptions of the numerical schemes can be found in these
papers. A speciality of these codes is the staggered grid: scalar
quantities (like density and pressure) are defined at the cell
centres, whereas the components of vector quantities (like velocity
and momentum) are defined at the cell walls. For the computation of
the gravitational potential an ADI solver is used
\citep[see][]{norman86}.  The code uses cylindrical coordinates
$(z,r)$ with the symmetry axis at $r=0$ and the inflow boundary at
$z=0$. 

The natural viscosity is neglected. However, the code makes use of an
artificial viscosity in order to handle shocks correctly. Without that
jump conditions and the propagation velocity of shocks could
not be reproduced correctly. The artificial viscosity is designed to
be non-zero in compressed regions only. It spreads shock fronts over a
few cells. Only the diagonal elements of the viscosity tensor
$\mathbf{Q}$ are non-zero. Due to the cylindrical symmetry also the
component $Q^{\phi\phi}$ is zero. The two remaining components take
the form
%
\bea
Q^{ZZ}(z,r) &=& \rho(z,r)\cdot \Delta_z v(z,r) \cdot \nonumber \\
&&[-C_1 c_s  +\, C_2 \cdot\min\{\Delta_z v(z,r), 0\}  ] ,  \\
Q^{RR}(z,r) &=& \rho(z,r)\cdot \Delta_r v(z,r) \cdot \nonumber \\
&&[-C_1 c_s +\, C_2 \cdot\min\{\Delta_r v(z,r), 0\}  ] ,
\eea
%
where $c_s$ is the adiabatic sound speed, $\rho(z,r)$ the local mass
density, $\Delta_z v(z,r)$ and $\Delta_r v(z,r)$ are the local
gradients of the velocity field in $z$- and $r$-direction,
respectively. The constant $C_1$ sets a linear term, which is,
according to \cite{norman86}, used rarely ``and then only sparingly to
damp oscillations in stagnant regions''. For the application here it
is also set to zero. The constant $C_2$ sets the quadratic term, it
gives roughly the number of cells over which a shock front is
spread. It is usually set to values between 1 and 4. Here we use
$C_2=2$ in agreement to previous work and in agreement with results
from numerical tests. A concern may be that due to the artificial
viscosity the code cannot handle hydrodynamical instabilities like the
Kelvin-Helmholtz instability correctly. We have tested the influence
of the viscosity in our simulations by varying the parameter $C_2$
(see Appendix~\ref{sec:viscosity}) and found that the differences are
negligible.

We do not yet include radiative cooling, because to do so properly we
would need to resolve the multiphase characteristic of the
interstellar medium (ISM), either by a multiphase code or by true
spatial resolution. However, this is not our aim here. In previous
works that included radiative cooling a lower limit for the ISM
temperature was used to prevent cooling to temperatures below $10^4$
or $10^5\K$, which prohibits to study the interesting temperature
range of the warm HI gas and also molecular
clouds. \citet{marcolini03} tested in how far including radiative
cooling leads to different results concerning the gas loss from the
galaxy and found no difference whether they include cooling or not.

The 2D treatment restricts the simulations to face-on winds, however,
\citet{marcolini03} showed by comparisons with 3D simulations that the
2D simulations can handle face-on winds correctly. What we can learn
for other inclinations from the pure face-on ones will be discussed in
Sect.~\ref{sec:discussion}.

In contrast to most applications of ZEUS-like codes for studies of a
flow past an object, we do not choose an open boundary condition for
the boundary at $r\Max$. Actually the conditions termed ``open
boundary'' only mean a semi-permeable boundary, where the outgoing
flow is extrapolated over the boundary, but no flow can enter.  For
consistency with the generally established expressions we continue to
use the term ``open boundary'' for the conditions just explained. We
found that for long run-times an open boundary at $r\Max$ leads to
slow ongoing loss of material through that boundary, which gradually
decreases the density of the wind. As this behaviour is not desirable,
we implemented a ``solid wall'' at $r\Max$, which means that the
velocity perpendicular to the boundary is set to zero, nothing can
cross the wall. The boundary at $z\Max$ is open in the sense explained
above.

\section{Initial Model}
\label{sec:initial}
%
\subsection{ICM}
%
In the beginning of the simulations the galaxy is embedded in a
homogeneous ICM at rest. ICM densities and temperatures cover the
range from the centres to the outskirts of clusters ($10^{-5}\ccm$ to
$10^{-2}\ccm$ for particle densities and $10^7\K$ to a few times
$10^7\K$ for temperatures). For further details see
Sect.~\ref{sec:sim_scan_wind}.
%
%
%
\subsection{Galaxy model}
\begin{figure*}
\includegraphics[height=0.5\textwidth,angle=-90]{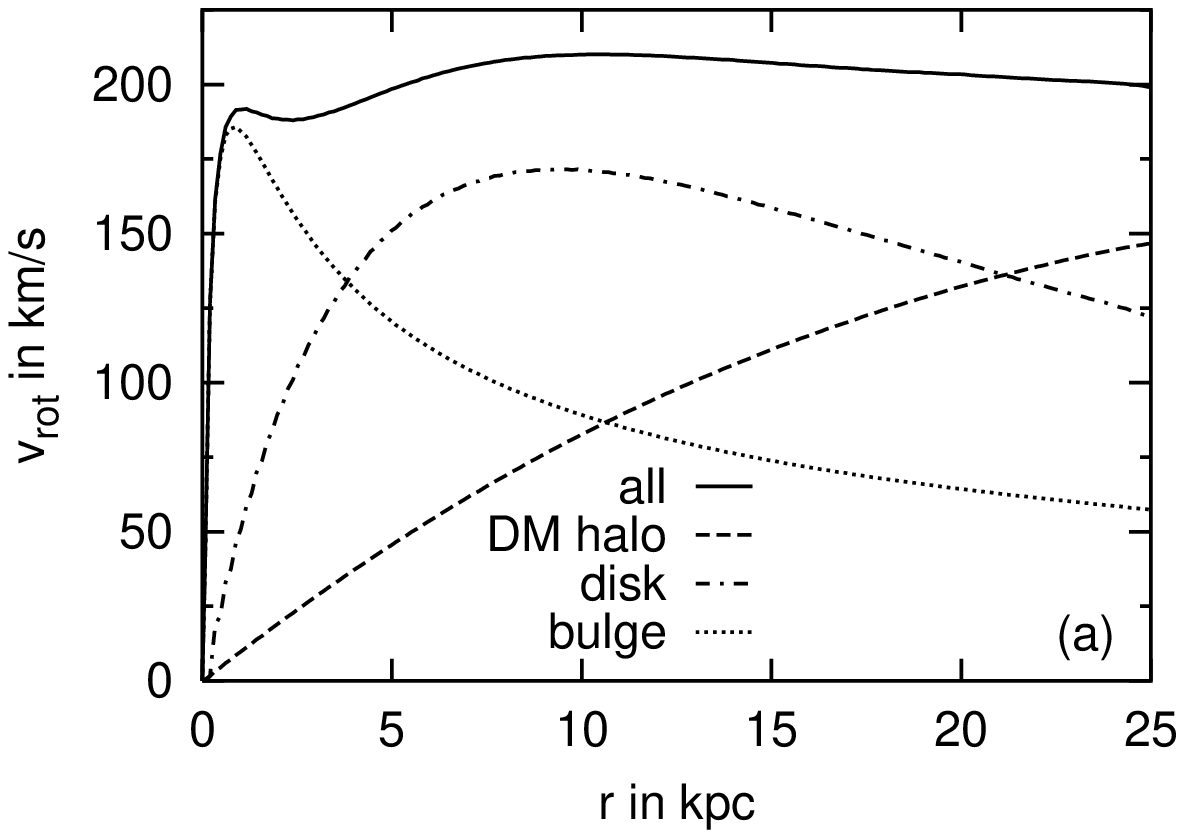}
\includegraphics[height=0.5\textwidth,angle=-90]{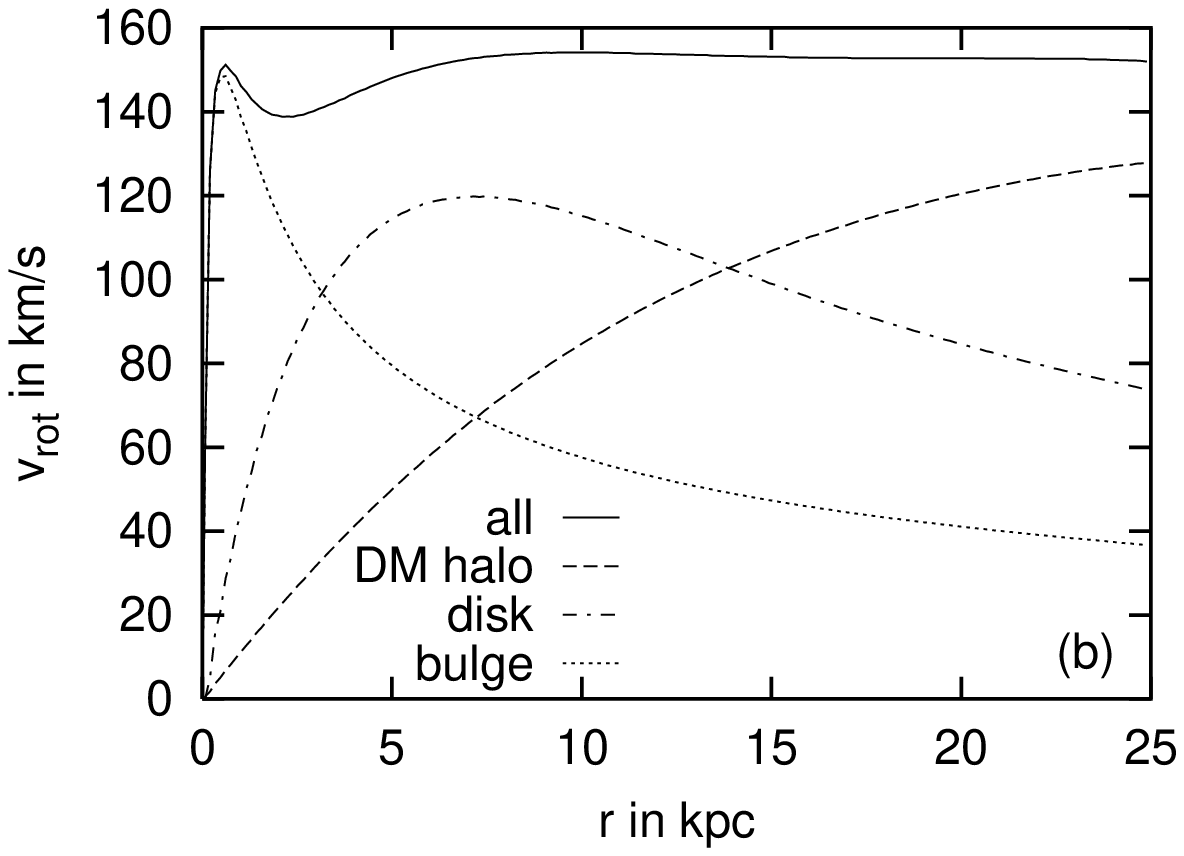}
\caption{Rotation curves and contributions of the single components for the massive (a) and the medium size galaxy (b).}
\label{fig:inimodel-vrot}
\end{figure*}

\begin{figure}[!tb]
\centering\resizebox{\hsize}{!}%
{\includegraphics[angle=-90]{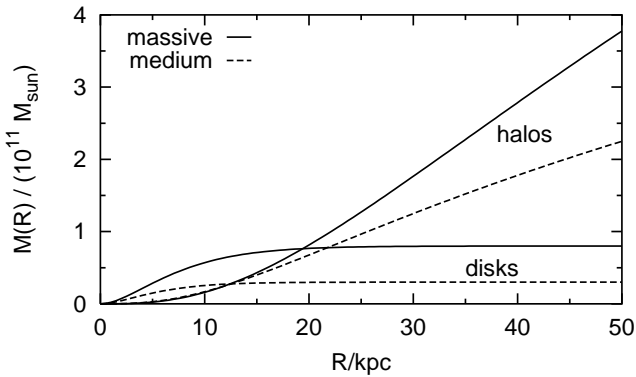}}
\caption{Cumulative masses (mass included in a sphere of radius $R$) for the stellar disk and the DM halo for the massive and the medium size galaxy.}
\label{fig:inimodel-cumulative-masses}
\end{figure}

%
We use two galaxy models, one resembling a massive galaxy with a
rotation velocity at the level of $200\,\Kms$, the other resembling a
medium mass galaxy with a rotation velocity of about $150\,\Kms$. Both
galaxies are composed of a dark matter (DM) halo, a stellar disk, a
stellar bulge and a gas disk. All components but the gas disk are
static, they provide the gravitational potential of the galaxy. The
specific model used for each component is introduced below, the
parameters for all components (masses and scale lengths) are
summarised in Table~\ref{tab:galaxy_parameters}.
\begin{table}
\caption{Model parameters for the massive (standard) and the medium mass galaxy. See also text for further explanations.}
\label{tab:galaxy_parameters}
\centering\begin{tabular}{llll} 
\hline\hline
&            &   massive               & medium \\
\hline
&$M\Stars$   & 8$\,\cdot 10^{10}M\Sun$ & 3$\,\cdot 10^{10}M\Sun$\\
stellar disk&$a\Stars$   & 4kpc                    & 3kpc \\
&$b\Stars$   & 0.25kpc                 & 0.25kpc \\
\hline
bulge &$M\Bulge$   & 2$\,\cdot 10^{10}M\Sun$ & 0.8$\,\cdot 10^{10}M\Sun$\\
&$R\Bulge$   & 0.4kpc                  & 0.15kpc \\ 
\hline
DM halo &$R\DM$      & 23kpc                               & 15kpc \\
\hline
&$M\Gas$     & $0.8\,\cdot 10^{10}M\Sun$ & $0.3\,\cdot 10^{10}M\Sun$\\
gas disk &$a_{\Sigma}$                    & 7kpc  & 7kpc\\
&$b\Gas ^{\mathrm{a}}$     & 0.4kpc                    & 0.4kpc  \\
&$v\Rot$     & 200$\Kms$                 & 150$\Kms$\\
\hline
\end{tabular} 
\begin{list}{}{}
\item[$^{\mathrm{a}}$] The value given is $b\Gas (r=11.5\Kpc)$. For 
the simple exponential disk this value is valid for all $r$; for the
flared disk $b(r)$ varies according to Eq.~\ref{eq:bgas_flared}.
\end{list}
\end{table}
Fig.~\ref{fig:inimodel-vrot} shows the decomposition of the rotation
curves into the contributions of the components for both the massive
and the medium mass galaxy.
Fig.~\ref{fig:inimodel-cumulative-masses} demonstrates the cumulative masses for the stellar disks and DM halos for both cases.

\subsubsection{Stellar disk}
The density of the stellar disk follows an exponential model:
\be
\rho(z,r)=\frac{M\Stars}{4\pi b\Stars a\Stars^2}  \exp\left(-\frac{r}{a\Stars}\right) \exp\left(-\frac{|z|}{b\Stars}\right) \label{eq:exp_disk}
\ee
The parameters are the total mass $M\Stars$, the scale radius
$a\Stars$ and the scale height $b\Stars$ (the asterisks mark the
parameters of the stellar disk). The potential needs to be computed
numerically.

\subsubsection{Stellar bulge}
The bulge is modelled following the suggestion of
\citet{hernquist93}. We assume a spherical bulge, which is described
by
\be
\rho\Bulge(R)=\frac{M\Bulge}{2\pi} \frac{R\Bulge}{R(R+R\Bulge)},
\ee
where $R\Bulge$ is the scale radius of the density distribution and
$M\Bulge$ the total mass.

\subsubsection{Dark matter halo}
For the DM halo the model by \citet{burkert95} is used. The potential
for this density distribution can be calculated analytically, it can
be found in \cite{mori00}.
\be
\rho (R) = \frac{\rho\DM R\DM ^3}{(R+R\DM)(R^2+R\DM ^2)},\label{eqn:burkert-DM-halo}
\ee
First this model was derived from rotation curves of dwarf disk
galaxies, but later on this form of DM halo could also be confirmed
for normal disk galaxies \citep{salucci00,borriello01}. According to
observations, the core radius $R\DM$, core mass $M_0$ (mass inside
$R\DM$) and central density $\rho\DM$ follow the scaling relations
\be
\rho\DM = 3\cdot 10^{-24} \left( \frac{R\DM}{\Kpc} \right)^{-2/3} \gccm ,
\ee
 so that Burkert-halos are essentially a one-parameter family.

\subsubsection{Gas disk} \label{sec:ini_gasdisk}
We use two different models for the gas disk. In general, for both the
density distribution can be written as
\be
\rho(z,r) = \frac{\tilde M\Gas}{4\pi b_0 a_\Sigma^2} \exp\left(-\frac{r}{a\Gas}\right) \exp\left(-\frac{|z|}{b\Gas(r)}\right) \label{eq:rho_gas}
\ee
and the surface density as
\be
\Sigma\Gas(r) = \frac{\tilde M\Gas}{2\pi a_\Sigma^2} \exp\left(-\frac{r}{a_\Sigma}\right),
\ee
where as usual $\tilde M\Gas$ is the total mass of the gas disk and
$a\Gas$, $b\Gas$ are the scale radius and the scale height,
respectively. The parameter $a_{\Sigma}$ is the scale radius for the
surface density.

For option one we choose $b\Gas(r)=b_0=\const$. This is again a usual
exponential disk like in Eq.~\ref{eq:exp_disk}. For this case also
$a\Gas$ and $a_\Sigma$ are identical.

The second option is a flared disk with
\be
b\Gas(r) = b_0 \exp\left(\frac{r}{r_b}\right). \label{eq:bgas_flared}
\ee
Here $a\Gas$ and $a_\Sigma$ are not identical, but they relate with
the parameter $r_b$ according to:
\be
\frac{1}{r_b} = \frac{1}{a\Gas} - \frac{1}{a_\Sigma}\quad\Leftrightarrow\quad 
r_b=\frac{a_\Sigma a\Gas}{a_\Sigma - a\Gas} .
\ee

To have an unperturbed flow past the galaxy, there must be enough
space between the outer edge of the gas disk and the $r\Max$ boundary
of the computational grid (we demand that the disk radius is smaller
than about 1/4 of $r\Max$). Due to the large $a_\Sigma$ the gas disk
would extend to large $r$, leaving not enough space between the disk
edge and the grid boundary, although the grid is already large (see
Sect.~\ref{sec:sim_grid}). To prevent such problems, we cut the gas
disk gradually to a finite radius of $26\,\Kpc$. If the disk gets
stripped at radii smaller than this, also the parts outside $26\,\Kpc$
would be stripped, so this does not bias the results. Due to this cut
the true mass of the gas disk $M\Gas$ is a bit lower than what is set
by the parameter $\tilde M\Gas$ in Eq.~\ref{eq:rho_gas}. The values
given in Table~\ref{tab:galaxy_parameters} are the true masses
$M\Gas$.

\begin{figure}
\centering
\resizebox{\hsize}{!}{\includegraphics[angle=-90]{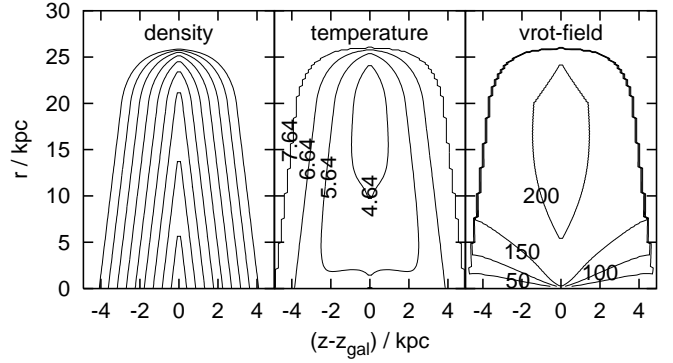}}
\caption{Contour plots for the density (left), temperature (middle) and 
rotation field (right) for the massive galaxy with a simple
exponential disk. Both the density and the temperature contours have a
logarithmic spacing. For the density plot there is one contour line
every half order of magnitude; the inmost contour at
$10^{-24}\gccm$. For the temperature the spacing of the contour lines
is one order of magnitude; the contours are labelled in units of
$\log_{10}(T/\K)$. In the velocity field the contours are labelled in
$\Kms$.  Please note that the disk appears thicker because the axes
have different scales.}
\label{fig:initial_contours_exp}%
\end{figure}

\begin{figure}
\centering
\resizebox{\hsize}{!}{\includegraphics[angle=-90]{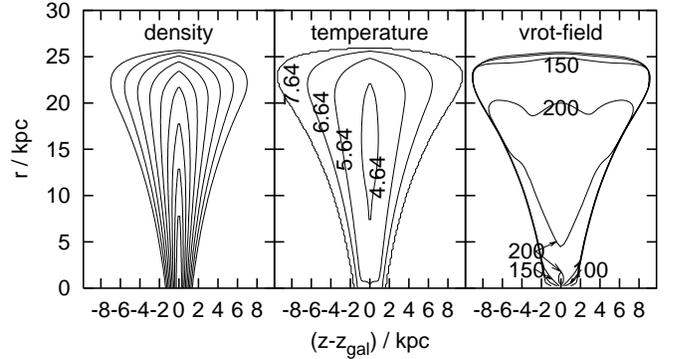}}
\caption{Same as Fig.~\ref{fig:initial_contours_exp} for the massive galaxy with a flared disk.}
\label{fig:initial_contours_flare}%
\end{figure}

\begin{figure}
\centering\resizebox{0.85\hsize}{!}{
\includegraphics[angle=0]{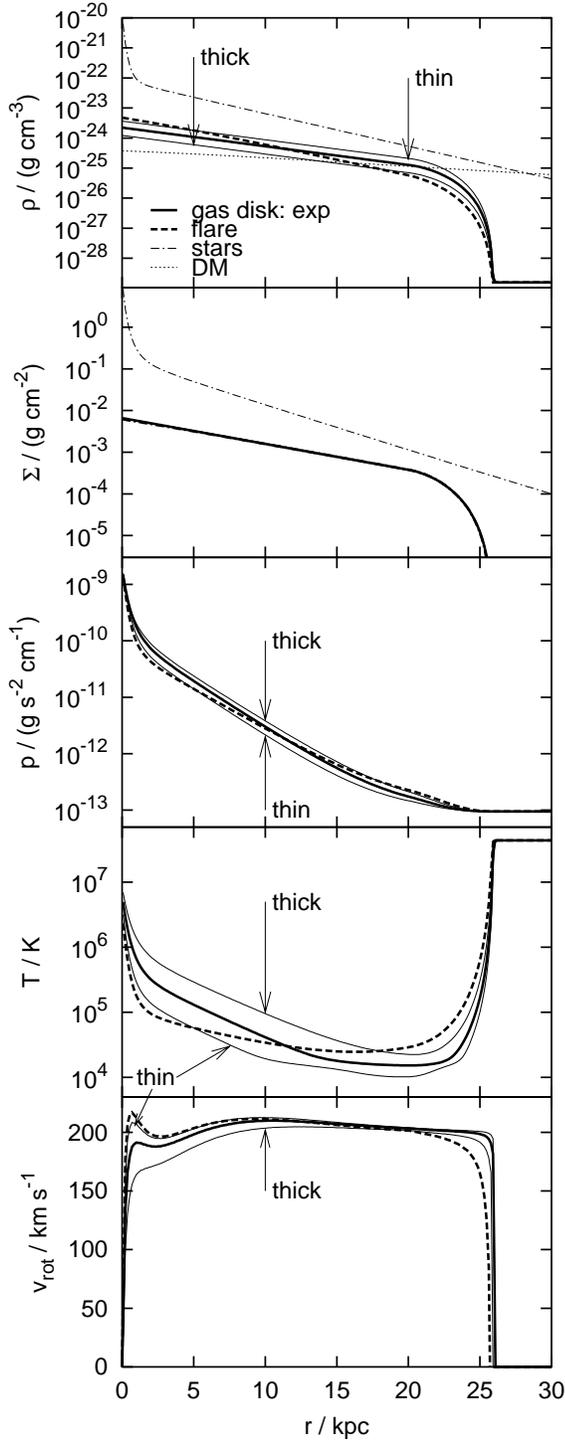}}
\caption{Profiles along the galactic plane for the massive galaxy. The 
panels show profiles for the density $\rho$, surface density $\Sigma$,
temperature $T$ and the rotation velocity $v\Rot$ of the gas disk; all
quantities are shown for the case of a normal exponential disk (thick
solid lines), thick and thin exponential disk (thin solid lines) and a
flared disk (thick dashed lines) (see
Sect.~\ref{sec:sim_thickness}). The surface density is the same for
all cases. We also show the density profiles for the stellar component
(disk+bulge) and the DM halo, and the surface density of the stellar
component. The thick lines correspond to the cases shown in
Figs.~\ref{fig:initial_contours_exp} and
\ref{fig:initial_contours_flare}.}
\label{fig:initial_profiles}%
\end{figure}

For stability reasons the gas disk needs to be in pressure equilibrium
with the surrounding ICM and in rotational equilibrium with the
galactic potential. Therefore, after setting all densities and
calculating the gravitational potential of the galaxy, we solve the
hydrostatic equation in $z$-direction. Thus knowing the pressure
distribution we compute the rotation velocity for the gas disk so that
in radial direction the gravitational force, the centrifugal force and
the force due to the radial pressure gradient are in equilibrium (the
pressure force is rather small in radial direction). The resulting
rotational velocities along the galactic plane for the massive and the
medium galaxy are shown in Fig.~\ref{fig:inimodel-vrot}. Density
contours, the temperature distribution and the rotation field for both
the normal exponential gas disk and the flared one can be seen in
Figs.~\ref{fig:initial_contours_exp} and
\ref{fig:initial_contours_flare}, respectively (both Figs.~are for the
massive galaxy).
%
%
%
In addition, Fig.~\ref{fig:initial_profiles} provides an overview of
profiles of the density, surface density, pressure and temperature
along the galactic plane.
%
%
Also the rotation curves are repeated as they differ slightly for the
exponential and the flared disk. Fig.~\ref{fig:initial_profiles} also
shows profiles for a thicker and a thinner (larger and smaller scale
height) exponential disk (see Sect.~\ref{sec:sim_thickness}).

\begin{figure}
\centering\resizebox{0.85\hsize}{!}{
\includegraphics[angle=-90]{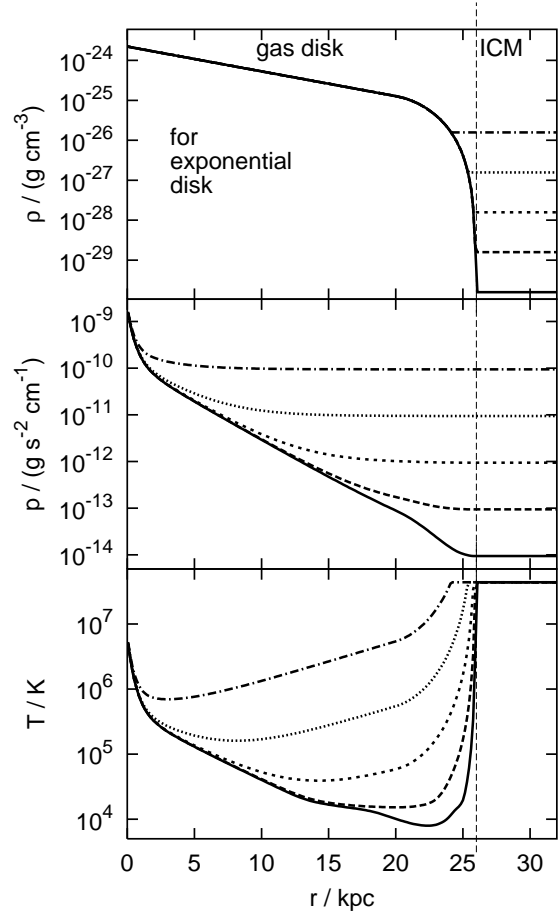}}
\caption{Profiles of the density $\rho$, pressure $p$ and temperature $T$ 
along the galactic plane, as they result from the hydrostatic
equilibrium for different ICM densities. The profiles are shown for
the case of an exponential gas disk. The density profile for the disk
was fixed, it does not change with changing ICM density. The ICM
densities, pressures and temperature can be read from the plot on the
rhs of the dashed vertical line (the ICM temperature of $4.385\cdot
10^7\K$ was the same in all cases). This plot demonstrates how
everywhere except in the very centre the pressure in the gas disk
adapts to the ICM pressure (see text for a discussion).}
\label{fig:comp_pres}%
\end{figure}

We want to stress that already the high external pressure has a strong
influence on the gas disk. If the galactic potential and the density
of the gas disk are given {\em and} the gas disk shall be in pressure
equilibrium with the surrounding ICM, then the pressure in the disk
increases with increasing ICM pressure in the way demonstrated in
Fig.~\ref{fig:comp_pres} for the exponential disk.
As explained above, for the given gas disk density and galactic
potential the pressure in the gas disk was computed assuming
hydrostatic equilibrium with the ICM in $z$-direction. The resulting
pressure and temperature profile inside $r=20\Kpc$ for the lowest ICM
pressure (solid lines in Fig.~\ref{fig:comp_pres}) are representative
for the case of an isolated galaxy (one not embedded in an ICM). From
the pressure profiles in Fig.~\ref{fig:comp_pres} it is already
obvious that e.g.~for an ICM with (particle) density of about
$10^{-3}\ccm \widehat{\approx} 10^{-27}\gccm$ and temperature
$T\ICM\approx 4\cdot 10^7\K$, which corresponds to conditions in the
cluster centre, its pressure is higher than the pressure in the disk
of an isolated galaxy at radii larger 10kpc. For this example, in case
of hydrostatic equilibrium with the ICM, the resulting pressure and
temperature profiles agree with the profiles in an isolated galaxy
only in the inner 3kpc. Outside this radius the pressure and
temperature of the disk are much higher than in an isolated
galaxy. However, as we want to start with a stable model, we use the
hydrostatic solutions as initial models. In how far this unrealistic
pressure and temperature profiles influence our results will be
discussed in Sect.~\ref{sec:discussion}.

\section{Simulations}
\label{sec:simulations}
The main purpose of this paper is to investigate how the mass loss
from the gas disk due to RPS depends on the strength of the ICM wind,
and on the galaxy itself. Therefore one needs to vary ICM parameters
as well as galaxy properties. A further critical point we check is how
the initialisation of the simulation influences the results. In this
chapter we list the models we have performed and explain our choices
of parameters. In addition, we explain some analysis techniques we
used to obtain the results.
%
%
\subsection{Grid size and Preparative Tests}  \label{sec:sim_grid}
%
The numerical grid must be large enough so that the boundaries do not
disturb the flow. After some tests we chose a grid size of
$100\,\Kpc\times100\,\Kpc$ and a resolution of $153\,\PC$ ($650^2$
grid cells). We tested the influence of the resolution by running some
representative cases with resolutions of $50\PC$, $100\,\PC$ and $200\,\PC$ and
found that the results are not sensitive to this (see Appendix~\ref{sec:resolution}). The galaxy is placed
at a distance of $z\Gal=40\Kpc$ from the inflow boundary. We also
tested that the exact position of the galaxy does not influence our
results as long as it is not placed too close to the boundaries. We
also checked that our results are not influenced by the assumption of
the static gravitational potential by performing a control run which
took the evolution of the potential due to the changing gas
distribution into account. We found that our results are not sensitive
to this, because the gas disk contributes only a minor fraction to the
total potential.
%
%
\subsection{Wind initialisation} \label{sec:sim_tswitch}
%
Initially all gas is at rest. To start the simulation we increase the
ICM velocity at the inflow boundary and the outflow boundary linearly
over the time interval $t\Switch$ from zero to the final value
$v\ICM$. Afterwards the inflow conditions are kept at the upstream
side boundary, and the downstream side boundary switches to the usual
open boundary condition (if $v\ICM$ is supersonic, the outflow
velocity at the downstream side boundary was only increased up to Mach
0.9 and then this boundary switched to the open condition). So
$t\Switch$ is the time during which the simulation is ``switched
on''. Two contradicting demands make it difficult to choose this
parameter. On behalf of a smooth flow $t\Switch$ should be as long as
possible. On the other hand, we expect that the galaxy responds to the
ram pressure on short time scales (a few $10\Myr$, see
Sect.~\ref{sec:analytic_instantaneous}). Therefore $t\Switch$ should
be rather short in order to resolve this time scale. As this choice is
not straightforward, we run some representative cases with $t\Switch
=20,70,150\Myr$. The shortest $t\Switch$ is chosen to resolve the
expected stripping time scale
(Sect.~\ref{sec:analytic_instantaneous}). The longest $t\Switch$ was
chosen to be as long as possible under the condition that the
initialisation must be finished before the first stripped material
reaches the outflow boundary. Stripped material that would reach the
outflow boundary during $t\Switch$ would cause reflection problems, as
during $t\Switch$ the outflow boundary has a fixed velocity and is not
open, as explained above. The maximum $t\Switch$ was found by trial
and error. The parameters for these runs are listed in
Table~\ref{tab:runs_tswitch}.
%
\begin{table}
\centering
\caption{Parameters for the simulations testing influence of $t\Switch$ 
(see Sect.~\ref{sec:sim_tswitch}). All these runs use the massive
galaxy with an exponential disk with $b\Gas=0.4\Kpc$ and $T\ICM{}_1$,
corresponding to a sound speed $c\ICM=1000\Kms$.}
\label{tab:runs_tswitch}
\begin{tabular}{cccc}
\hline\hline
$\tilde p\Ram$   & $v\ICM$  & $n\ICM$             & $t\Switch$ \\
$(\Rampresunit)$ & $(\Kms)$ & $(\ccm)$            & $(\Myr)$      \\
\hline
1000             & 800      & $1.56\cdot 10^{-3}$ & 20, 70, 150 \\ 
1000             & 2530     & $1.56\cdot 10^{-4}$ & 20, 70, 150 \\ 
100              & 800      & $1.56\cdot 10^{-4}$ & 20, 70, 150 \\ 
100              & 2530     & $1.56\cdot 10^{-5}$ & 20, 70, 150 \\ 
\hline
\end{tabular} 
\end{table}
%
%
\subsection{Wind parameters} \label{sec:sim_scan_wind}
%
To investigate the influence of the wind strength we expose the
massive galaxy with a standard gas disk (exponential, $b\Gas=0.4\Kpc$)
to different ICM winds. These winds resemble conditions reaching from
the centres to the outskirts of galaxy clusters.

Typical velocities of cluster galaxies (and hence the wind velocities)
are of the same order as the sound speed of the ICM. The ICM
temperatures in clusters are between $10^7\,\K$ and $10^8\,\K$. For
the standard ICM temperature we choose $T\ICM{}_1=4.385\cdot
10^7\,\K$. With a mean molecular weight of $\mu=0.6$ for the ionised
ICM this temperature corresponds to a sound speed of $1000\,\Kms$. For
some simulations a lower ICM temperature of
$T\ICM{}_2=T\ICM{}_1/\sqrt{10}=1.39\cdot 10^7\,\K$ corresponding to a
sound speed of $1000\,\Kms/\sqrt[4]{10}=562\,\Kms$ is used. If not
stated otherwise, we use $T\ICM{}_1$. ICM particle densities $n\ICM$
range from $10^{-5}$ to a few $10^{-3}\ccm$. We additionally include
extreme cases for the density to be able to scan the ICM parameter
space systematically.

\begin{figure}[!tb]
\resizebox{\hsize}{!}%
{\includegraphics[angle=-90]{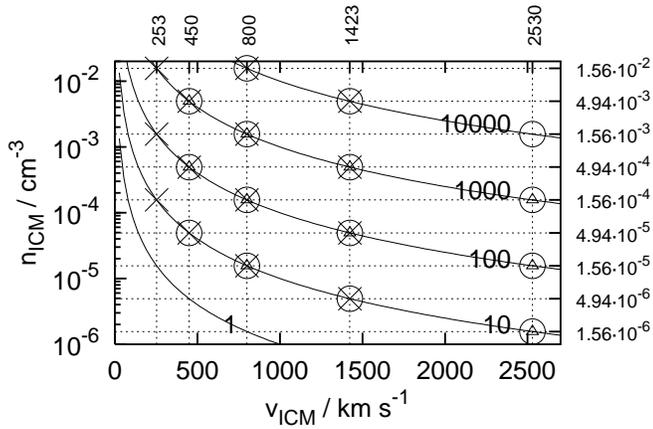}}
\caption{Models in the $v\ICM$-$n\ICM$ plane. Contours of constant specific 
ram pressure $\tilde p\Ram$ are plotted with a spacing of one order of
magnitude. The lines are labelled with $\tilde p\Ram$ in units of
$\Rampresunit$. The open circles mark the $(v\ICM,n\ICM)$ pairs used
for the simulations with the massive galaxy and $T\ICM{}_1$. The
triangles mark the winds used for simulations of the medium mass
galaxy and the same $T\ICM$. The crosses mark the winds used for the
massive galaxy and $T\ICM{}_2$ (see also
Sect.~\ref{sec:sim_scan_wind}).}
\label{fig:parameterspace_wind}
\end{figure}

For the sake of suitable numbers we will use the specific ram pressure
\be
\tilde p\Ram = n\ICM v\ICM^2
\ee
in units of $\Rampresunit$ throughout the paper. It relates to the real 
ram pressure $p\Ram$ through
\bea
p\Ram &=& \mu\cdot\mathrm{amu}\cdot \tilde p\Ram \nonumber \\
      &=& 1.002\cdot 10^{-14} \frac{\tilde p\Ram}{\Rampresunit} \;\Erg\,\ccm .
\eea
In Fig.~\ref{fig:parameterspace_wind} we show isolines of $\tilde p\Ram$ in t
he $(v\ICM,n\ICM)$ plane. 
%
Our simulations cover a range from 10 to $10\,000\,\Rampresunit$ in
$\tilde p\Ram$ ($10^{-13}$ to $10^{-10}\Erg\,\ccm$ in $p\Ram$).

We choose pairs of $(v\ICM,n\ICM)$ so that every $\tilde p\Ram$ is
covered by four different combinations of $v\ICM$ and $n\ICM$, two in
the subsonic and two in the supersonic regime. In this way we can see
if the success of ram pressure stripping depends on $\tilde p\Ram$
alone or if it also depends on the Mach number. The pairs chosen for
the simulations with the massive galaxy and $T\ICM{}_1$ are marked by
open circles in Fig.~\ref{fig:parameterspace_wind}.

To study of the influence of the Mach number in more detail, the
simulations are repeated with $T\ICM{}_2$ (marked by crosses in
Fig.~\ref{fig:parameterspace_wind}). We choose $T\ICM$, $v\ICM$ and
$n\ICM$ so that the velocities $v\ICM=253,450,800,1423,2530\,\Kms$
correspond to Mach numbers $0.253,0.45,0.8,1.423,2.53$ for
$T\ICM{}_1$, and to Mach numbers $0.8,1.423,2.53,4.5$ for $T\ICM{}_2$,
respectively. In this way many cross-comparisons are
possible. E.g.~the stripping effect of a wind with a velocity of
$800\,\Kms$ can be studied once for the case this velocity corresponds
to Mach 0.8 and once to Mach 1.423.
%
%
\subsection{Influence of the galactic parameters}
%
In this paper we focus on two aspects of the galaxy. First we address
the question whether the stripping efficiency is determined solely by
the gas surface density as suggested by the Gunn \& Gott criterion
(Eq.~\ref{eq:gunngott}), or if the vertical structure of the gas disk
plays a role. The second point of interest is the influence of the
total mass of the galaxy.
%
%
\subsubsection{Influence of the vertical structure of the gas disk} \label{sec:sim_thickness}
To study the influence of the shape and thickness of the gas disk on
the RPS efficiency we expose exponential and flared gas disks with
different $b\Gas$ to some representative ICM winds. The simulations
done for this point of interest are listed in
Table~\ref{tab:runs_thickness}.
%
\begin{table}
\centering
\caption{Parameters for the simulations testing influence of the vertical structure of the gas disk (see Sect.~\ref{sec:sim_thickness}). Common to all these runs are the massive galaxy, $T\ICM{}_1$, $t\Switch=20\Myr$. The ``$\cdot$'' means the same value as in the previous line.}
\label{tab:runs_thickness}
\begin{tabular}{ccccc}
\hline\hline
$\tilde p\Ram^{\mathrm{a}}$  & $v\ICM^{\mathrm{a}}$ & $n\ICM^{\mathrm{a}}$           &gas disk&$b\Gas^{\mathrm{b}}/\Kpc$ \\
\hline
1000           & 800     &$1.56\cdot 10^{-3}$&  exp   &  0.2, 0.4, 0.8      \\
$\cdot$        & $\cdot$ & $\cdot$           & flared &  0.2, 0.4            \\
1000           & 2530    &$1.56\cdot 10^{-4}$&  exp   &  0.2, 0.4, 0.8       \\
$\cdot$        & $\cdot$ & $\cdot$           & flared &  0.2            \\
100            & 800     &$1.56\cdot 10^{-4}$&  exp   &  0.2, 0.4, 0.8      \\
$\cdot$        & $\cdot$ & $\cdot$           & flared &  0.2            \\
100            & 2530    &$1.56\cdot 10^{-5}$&  exp   &  0.2, 0.4, 0.8     \\
$\cdot$        & $\cdot$ & $\cdot$           & flared &  0.2            \\
\hline 
\end{tabular} 
\begin{list}{}{}
\item[$^{\mathrm{a}}$] The units of $\tilde p\Ram$, $v\ICM$ and $n\ICM$ are the same as in Table~\ref{tab:runs_tswitch}.
\item[$^{\mathrm{b}}$] The value given is $b(r=5\Kpc)$. For the simple exponential disk this value is valid for all $r$; for the flared disk $b(r)$ changes according to Eq.~\ref{eq:bgas_flared}.
\end{list}
\end{table}
%
%
\subsubsection{Influence of galaxy mass}
In order to check our results for a smaller galaxy (see right column
of Table~\ref{tab:galaxy_parameters}) we rerun the simulations with
different wind parameters using $T\ICM{}_1$. The wind velocities and
particle densities used here are marked in
Fig.~\ref{fig:parameterspace_wind} by small open triangles.
%
%
%
%
\subsection{Analysis techniques -- gas disk mass and radius}
%
For all simulations we need to calculate the mass and radius of the
remaining gas disk. Concerning the fate of the stripped gas, with our
code we can trace how much of the stripped gas falls back to the disk
and how much is lost permanently.
%
%
\subsubsection{Definitions} \label{sec:definitions}
We use the following definitions:
\begin{description}
\item[\textbf{The original disk region}] is defined as the region where 
     $\rho>\rho\ICM$ at the start of the simulation.
\item[\textbf{Galactic gas}] is gas that has been inside the original 
     disk region initially. The method used to trace this gas is
     described in Sect.~\ref{sec:colouring}.
\end{description}

We measure at each time $t$:
\begin{itemize}
\item[(a)] mass $M\Origreg(t)$ of galactic gas inside the original disk region,
\item[(b)] mass $M\Cylreg(t)$ of galactic gas inside a fixed cylinder 
           centred on the galactic plane with $r<26\,\Kpc$ and
           $|z-z\Gal|<5\,\Kpc$,
\item[(c)] mass $M\Bound(t)$ of galactic gas bound to the galaxy potential. 
           Bound gas is identified by a negative total energy density
           $e\Tot = e\Pot + e\Therm + e\Rot + e\Kin$ (potential,
           thermal, rotational and kinetic energy density).
\item[(d)] mass $M\Fallback(t)$ of galactic gas that has left the original 
           disk region at some moment but is now back in the original
           disk region -- that is the amount of fallen back gas (see
           Sect.~\ref{sec:colouring}).
\end{itemize}

The radius of the remaining disk $r\Disk(t)$ is determined by scanning
along the disk plane, starting at $r=0$. The first cell in which the
density drops below $10^{-26}\gccm$ defines the radius of the gas disk
(see Fig.~\ref{fig:denscont_thickness}). This seems rather arbitrary,
but we tested several other versions, and this one turned out to be
the most appropriate and representative one.
%
%
\subsubsection{Colouring technique -- tracing the galactic gas} \label{sec:colouring}
To follow the fate of the galactic gas we use the technique of
``colouring'' the gas. For its basic version
\citep{severing95,lohmann00} one has to do the following: Define an
additional array COLOUR1. At the start of the simulation, copy the
density of the galactic gas to this array. During the simulation
advect this array in the same way as the usual density. Thus the value
of COLOUR1 in a certain cell tells the density of galactic gas at this
position.

We extended this method to determine how much of the stripped galactic
gas falls back to the original disk region. Therefore gas of colour 1
(galactic gas) that leaves the original disk region ``changes its
colour'' to colour 2. This is done with the help of a second array
COLOUR2, which is set to zero initially. Then at each time step the
procedure is as follows: for each grid cell outside the original disk
region add the value of COLOUR1 to the value of COLOUR2 in the same
cell, then set COLOUR1 to zero in this cell. This causes the stripped
gas to change its colour. Stripped gas with colour 2 that comes back
to the disk region does not change its colour again, but keeps colour
2. Hence integrating over the array COLOUR2 inside the original disk
region gives the amount of galactic gas that has been outside at some
time -- which is the amount of galactic gas that has been stripped and
fallen back $M\Fallback$. Integrating over COLOUR1 gives the amount of
galactic gas that has never been outside the original disk region
$M\Stay$, and $M\Fallback+M\Stay=M\Origreg$ is the total amount of
galactic gas inside the original disk region, no matter if it has been
outside this region in the meantime or not.

\section{Results}
\label{sec:results}
In this Sect. we first make some analytical estimates and explain what
we expect to happen. After that we present the results of the
simulations and discuss them.
%
%
\subsection{Analytical Estimates} \label{sec:analytic}
%
As discussed by previous authors \citep[e.g.][]{mori00,marcolini03},
the ICM-ISM interaction can be divided into two sub-processes:
instantaneous stripping and continuous stripping. The first process is
determined by the force balance between the ram pressure and the
gravitational resistance. Depending on the strength of the ram
pressure, the outer gas disk down to a radius $r\Strip$ should be
pushed away in a relatively short time (see
Sect.~\ref{sec:analytic_instantaneous}). In contrast, the continuous
(turbulent viscous) stripping \citep[see][]{nulsen82} is a phenomenon
connected with the boundary between the ICM and the gas disk and hence
works on the whole surface of the gas disk. It has a rather long time
scale (see Sect.~\ref{sec:analytic_continuous}). Due to the different
time scales of the two subprocesses the overall process should appear
to happen in two phases, first the instantaneous stripping, followed
by the continuous stripping phase.

\subsubsection{Instantaneous stripping} \label{sec:analytic_instantaneous}
\begin{figure}
\centering\resizebox{\hsize}{!}{\includegraphics[angle=-90]{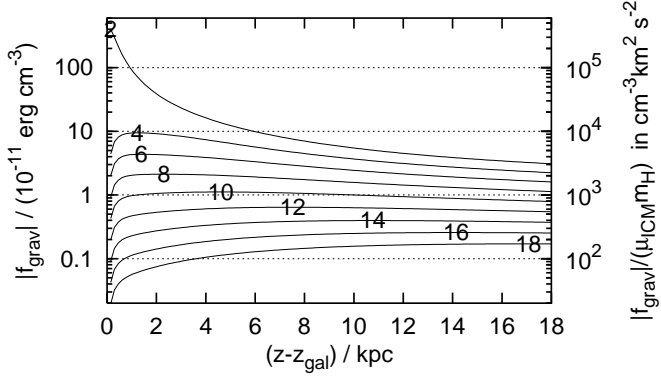}}
\caption{Profiles of the gravitational restoring force per unit area in 
$z$-direction $f\Grav(z,r)$ behind the disk, at different radii
$r$. Each profile is labelled with its cutting radius in kpc. The
label is placed at the maximum of the particular profile. Please note
that for larger $r$ the strongest gravitational resistance is far
behind the disk. For comparison, the right y-axis is labelled in units
of $\Rampresunit$ (unit of $\tilde p\Ram$).}
\label{fig:fgrav}%
\end{figure}

As mentioned in the introduction, \citet{gunn72} suggested to compare
the ram pressure and the gravitational restoring force per unit area
to estimate which part of the gas disk will be pushed out by the ICM
wind. We consider this estimate a bit closer here. A crucial point is
that it assumes that the gas disk is thin enough that its mass
distribution can be represented solely by its surface density
$\Sigma\Gas(r)$. Secondly, the gravitational acceleration in
$z$-direction in any position is $\frac{\partial\Phi}{\partial
z}(z,r)$. So, strictly speaking,
\be
|f\Grav(z,r)| = \left|\frac{\partial\Phi}{\partial z} (z,r) \Sigma\Gas(r)\right|, \label{eq:fgrav}
\ee
is the gravitational force per unit area in $z$-direction that works
on the gas disk at radius $r$ if this gas disk is shifted to position
$z$ in the galactic potential. We point this out in such detail,
because for a given $r$, $f\Grav(z,r)$ changes with $z$. E.g.~in the
most simple case, at $z=z\Gal$ the acceleration due to the
gravitational potential $\frac{\partial\Phi}{\partial z}=0$ and so is
the restoring force. In Fig.~\ref{fig:fgrav} we plot profiles of
$f\Grav(z,r)$ perpendicular to the galactic plane (in $z$-direction),
on the downstream side of the galaxy, for different radii.
As expected, for small radii the restoring force is
strongest. Interestingly, with increasing $r$ the maximum of
$f\Grav(z)$ is located at increasingly large distances from the
galactic plane. So for radii $>10\Kpc$ the maximum of $f\Grav(z)$ is
{\em outside} the initial gas disk \citep[this feature was already
mentioned briefly by][]{schulz01}. Which $f\Grav(z)$ should be used
for the estimation of $r\Strip$? Two plausible choices are possible:
either to use the maximal $f\Grav(z)$ {\em inside} the original gas
disk, or to use the maximal $f\Grav(z)$ at any $z$. For the comparison
of $p\Ram$ and $f\Grav(r)$ the two version can lead to different
results concerning radii $\ga 10\Kpc$.

\begin{figure}
\resizebox{\hsize}{!}{\includegraphics[height=0.4\textwidth,angle=-90]{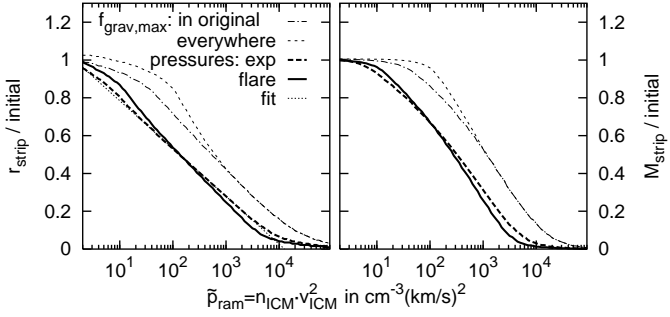}}
\caption{Analytical estimate of stripping radius $r\Strip$ and mass 
$M\Strip$ of the remaining gas disk after the instantaneous stripping
as a function of $\tilde p\Ram$. The thin dash-dotted and short-dashed
lines are derived from comparing the gravitational restoring force
$f\Grav$ with the ram pressure, the dash-dotted line uses the maximum
of $f\Grav(z)$ inside the original gas disk, the short-dashed line is
the estimate using $f\Grav(z)$ at any $z$. The solid and long-dashed
thick lines are derived from comparing the thermal pressure in the
galactic plane with the ram pressure, for the flared and the
exponential disk, respectively. See also Sect.~\ref{sec:analytic}. The
dotted line is the fit to the result of the pressure comparison for
the exponential disk according to Eq.~\ref{eq:fit_Rstrip}.}
\label{fig:analytic}%
\end{figure}

At the stripping radius $r\Strip$, $p\Ram$ and $f\Grav(r)$ are
equal. For smaller $r$ the gas is retained in the galaxy as
$p\Ram<f\Grav(r)$, for larger $r$ the gas will be lost because
$p\Ram>f\Grav(r)$. The resulting $r\Strip$ as a function of $\tilde
p\Ram$ is shown in the left panel of Fig.~\ref{fig:analytic} for the
massive galaxy for both versions of $f\Grav(z)$.
The right panel shows the mass of the gas disk remaining after the
instantaneous stripping $M\Strip$, i.e.~all gas at radii smaller than
$r\Strip$. The estimates according to the two versions differ only for
lower $\tilde p\Ram$, and the difference is small.

We note that for gas disks with larger scale heights the assumption
that the mass distribution is represented well enough by the surface
density alone may be wrong. In this case the gravitational restoring
force per unit area on a gas package of volume $\Delta V$ is
\bea
|f\Grav(z,r)| &=& \left|\frac{\partial\Phi}{\partial z} (z,r) \frac{\rho\Gas(r) \Delta V}{\Delta A} \right| \nonumber \\
&=& \left|\frac{\partial\Phi}{\partial z} (z,r) \rho\Gas(r) \Delta z \right|, \label{eq:fgrav2}
\eea
where $\Delta A$ the surface of the gas package perpendicular to the
wind direction and $\Delta z$ is the length of this package in wind
direction. To obtain an upper limit, for $\rho\Gas(r)$ the density in
the galactic plane is used. It is not obvious which value should be
used for $\Delta z$, so that this is rather a free parameter (which
should be somewhere between the grid resolution and the overall
thickness of the disk). If the scale height $b\Gas$ is the same for
all radii $r$, the expression in Eq.~\ref{eq:fgrav2} is proportional
to the one in Eq.~\ref{eq:fgrav}, but smaller; Eq.~\ref{eq:fgrav}
would overestimate the restoring force. With $f\Grav$ from
Eq.~\ref{eq:fgrav2}, the predicted $r\Strip(\tilde p\Ram)$ has the
same shape as before (using Eq.~\ref{eq:fgrav}), but is shifted in
horizontal direction towards lower $\tilde p\Ram$. If however the gas
disk is flared, the expressions in Eqs.~\ref{eq:fgrav} and
\ref{eq:fgrav2} are not proportional and the analytical estimates of
$r\Strip(\tilde p\Ram)$ would have different shapes.

\citet{mori00} suggest an alternative criterion for the stripping 
efficiency for spherical galaxies, namely a spherical galaxy will be
stripped completely if the ram pressure exceeds the thermal pressure
in the galactic centre. We modify this estimate for the application to
disk galaxies. Given the radial thermal pressure profile $p_0(r)$ in
the galactic plane of an {\em isolated} galaxy (as shown in
Fig.~\ref{fig:comp_pres}), the gas will be stripped from radii $r$
where
\be
p\Ram > p_0(r). \label{eq:estimate_comp_pres}
\ee
Again, $r\Strip$ is the radius where $p\Ram = p(r)$. The result from
this estimate is also shown in Fig.~\ref{fig:analytic} for both the
exponential and the flared disk. As the pressure profiles for both
cases are similar, also the functions $r\Strip(\tilde p\Ram)$ are
similar. The difference between the pressure profiles for the
different disk scale heights (thick and thin disks) is not larger than
the difference between the pressure profiles for the flared and the
exponential disk (see Fig.~\ref{fig:initial_profiles}). Hence the
profile for the medium scale height is representative for all
thicknesses.  Interestingly, the shape of the function $r\Strip(\tilde
p\Ram)$ is very similar to the shape of the estimate from the Gunn \&
Gott criterion, but the result from the pressure comparison is offset
towards smaller ram pressures, indicating that
$p_0(r)<f\Grav{}\Max(r)$. This can be understood analytically. The
pressure in the galactic plane at a certain radius $p_0(r)$ derived
from hydrostatics (for an isolated galaxy, i.e.~$p(\pm\infty)=0$) is
\be
p_0(r) = \int\limits_{-\infty}^{z\Gal}  -\frac{\partial\Phi}{\partial z}(z) \,\rho(z)\,\mathrm{d}z.  \label{eq:p0}
\ee
The calculation of $|f\Grav|$ uses the steepest potential gradient
$\left|\frac{\partial\Phi}{\partial z} (r) \right |\Max$ and the
surface density
\be
\Sigma(r)=\int\limits_{-\infty}^{\infty} \rho(z)\mathrm{d}z = 2\int\limits_{-\infty}^{z\Gal} \rho(z)\mathrm{d}z.
\ee
With that we can write $|f\Grav(r)|$ as
\be
|f\Grav(r)| = \int\limits_{-\infty}^{z\Gal}  2\left|\frac{\partial\Phi}{\partial z} (r) \right |\Max \rho(z) \mathrm{d}z.  \label{eq:fgrav_detail}
\ee
As in Eqs.~\ref{eq:p0} and \ref{eq:fgrav_detail} all functions are
positive in the integral range and as
$\left|\frac{\partial\Phi}{\partial z} \right |\Max \ge
-\frac{\partial\Phi}{\partial z}$ in this range, it follows that
\be
|f\Grav(r)| \ge 2 p_0(r).
\ee
For our galaxy model we have $f\Grav(r)\approx 4p_0(r)$. 

The time scale for the instantaneous stripping $t\Strip$ can be
estimated from the simple uniformly accelerated movement. With a
constant acceleration $a\approx p\Ram/\Sigma\Gas$ a body that starts
with zero velocity moves the distance $s$ during the time
\bea
t\Strip &=& \sqrt{\frac{2s}{a}} = \sqrt{\frac{2s\Sigma\Gas}{p\Ram}} \\
        &=& 56 \left(\frac{s}{5\Kpc}\right)^{1/2} 
               \left(\frac{\Sigma\Gas}{10^{-3}\mathrm{g}\,\Cm^{-2}}\right)^{1/2} \nonumber\\ 
        &&\times\left(\frac{\tilde p\Ram}{1000\Rampresunit}\right)^{-1/2} \Myr.
 \label{eq:t_strip}
\eea
Let us assume that a gas package is stripped off when it has reached a
distance of 5kpc to the disk plane. For the medium galaxy the surface
density is about a factor of 0.38 lower, hence the corresponding
stripping time scale for the medium galaxy should be about a factor of
$\sqrt{0.38}\approx 0.6$ shorter. This estimate is less accurate for
small ram pressures because in that case also the gravitational
restoring force needs to be taken into account (for small $p\Ram$ this
estimate gives a lower limit). For large ram pressures the
gravitational deceleration can be neglected because it is small
compared to the acceleration due to $p\Ram$.

\subsubsection{Continuous stripping}  \label{sec:analytic_continuous}
\begin{figure}
\centering\resizebox{0.7\hsize}{!}{\includegraphics[angle=-90]{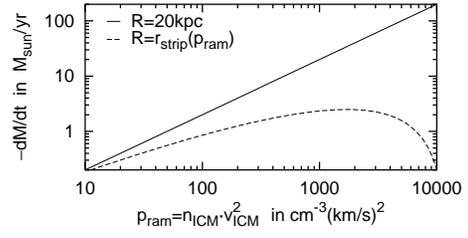}}
\caption{Analytical estimate of the mass loss rate for the continuous 
stripping. The solid line shows the result if a constant radius for
the gas disk is assumed, the dashed line shows the reduced mass loss
rate if $r\Strip$ as a function of $\tilde p\Ram$ is used
(Eq.~\ref{eq:fit_Rstrip}). Both functions are shown for
$v\ICM=1000\Kms$.  One has to keep in mind that this estimate is only
rough, as it is derived for a spherical and not a disk galaxy.}
\label{fig:analyt_KHmasslossrate}%
\end{figure}

According to \citet{nulsen82} the maximal mass loss rate of a
spherical cloud of radius $R$ in an ICM wind with particle density
$n\ICM$ and wind velocity $v\ICM$ is
\be
\dot M \approx 20 \left(\frac{R}{20\Kpc}\right)^2 \left(\frac{n\ICM}{10^{-3}\ccm}\right) \left(\frac{v\ICM}{1000\Kms}\right) \frac{M\Sun}{\Yr}. \label{eq:nulsen_pure}
\ee
Although this estimate was derived for spheres, also a rough estimate
for disks can be deduced. Apart from the geometry, for the face-on
wind flowing past a disk galaxy, the radius of the remaining gas disk
depends on the strength of the wind, as the gas disk is truncated
quickly by the instantaneous stripping. Over a wide range of ram
pressures the stripping radius can be fitted by
\bea
\frac{r\Strip}{r_0}&=&1.037-0.253\log_{10}\left(\frac{\tilde p\Ram}{\Rampresunit}\right) \nonumber \\
 &&\textrm{for the initial disk radius}\quad r_0=24.54\Kpc \quad\textrm{or} \nonumber \\
\frac{r\Strip}{\Kpc} &=& 25.44-6.2\log_{10}\left(\frac{\tilde p\Ram}{\Rampresunit}\right), \label{eq:fit_Rstrip}
\eea
as is shown in Fig.~\ref{fig:analytic}. We can rearrange Eq.~\ref{eq:nulsen_pure} to
\bea
\dot M &\approx& 20 \left(\frac{R}{20\Kpc}\right)^2 \times \nonumber \\
&& \left(\frac{\tilde p\Ram}{10^3\Rampresunit}\right) \left(\frac{v\ICM}{10^3\Kms}\right)^{-1} \frac{M\Sun}{\Yr} \label{eq:nulsen_new}
\eea
and replace $R$ by $r\Strip(\tilde p\Ram)$, thus deriving a rough estimate for the mass loss rate for our case. Fig.~\ref{fig:analyt_KHmasslossrate} compares the predicted mass loss rates of the continuous stripping as a function of $\tilde p\Ram$ for $v\ICM=1000\Kms$ for a constant $R$ on the one hand, and for $R=r\Strip=r\Strip(\tilde p\Ram)$ on the other hand.
If we take into account that the radius is truncated by the
instantaneous stripping, the mass loss rate is reduced
significantly. However, this estimate can only be rough as it was
derived for spherical bodies instead of disks. Beyond that the
dependence on $v\ICM$ (see Eq.~\ref{eq:nulsen_new}) introduces a
further factor. So we can only estimate that the mass loss rate due to
the continuous stripping is of the order of one
$M\Sun\,\Yr^{-1}$. With this number the time scale for the continuous
stripping
\be
t\Visc = \frac{M\Gas(t=0)}{\dot M}
\ee
is of the order of several Gyr.
%
%
\subsection{General behaviour of the simulations} \label{sec:result_general}
\begin{figure}
\resizebox{\hsize}{!}{\includegraphics[angle=-90]{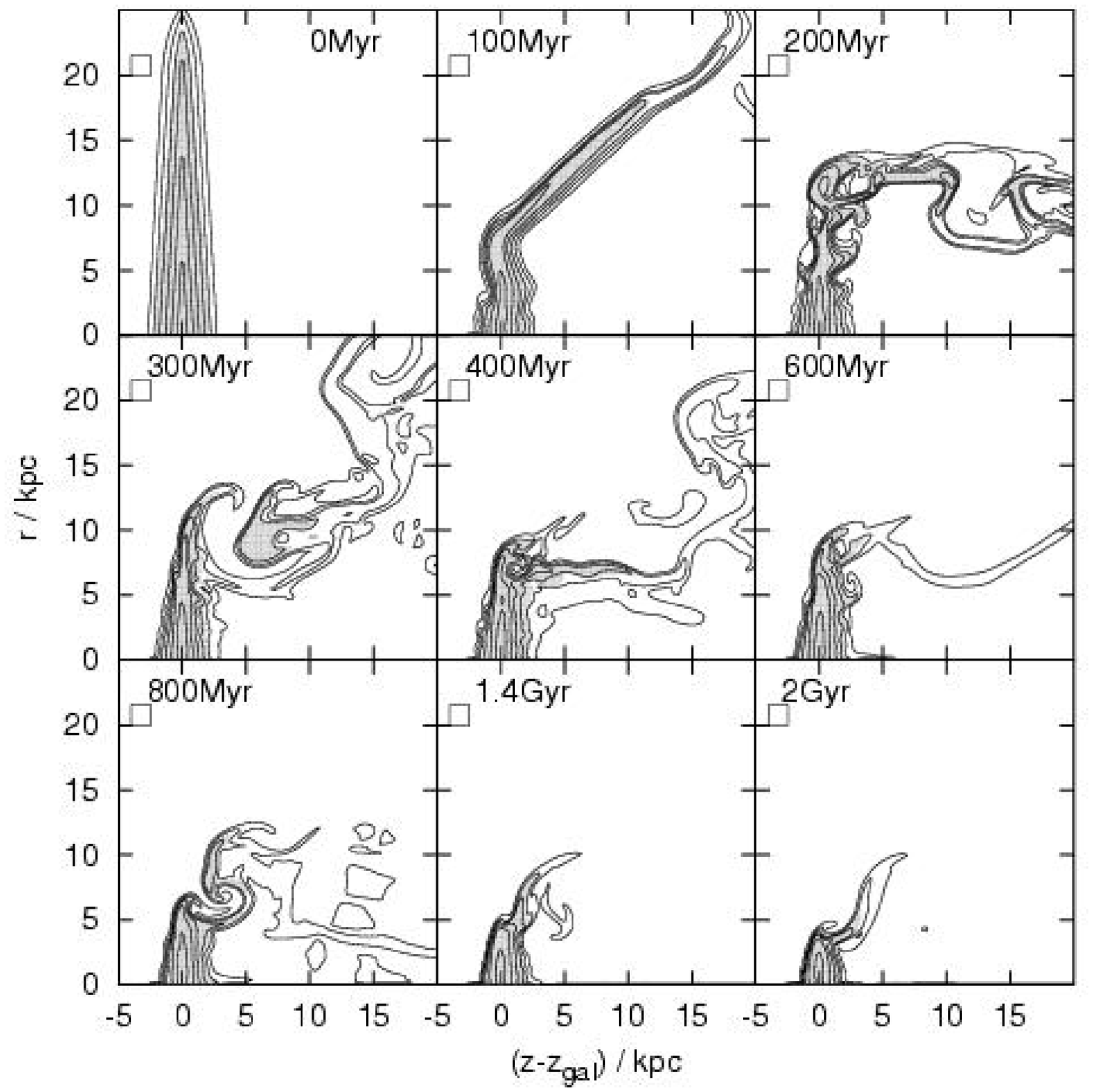}}
\caption{Evolution of gas density distribution for the massive galaxy in 
a wind with $T\ICM{}_1$, $\tilde p\Ram=1000\Rampresunit$, Mach 0.8,
$t\Switch=20\Myr$. The density distribution is shown by contour lines
with the inmost contour at $10^{-24}\gccm$; the spacing of the
contours is half an order of magnitude. Only contours for
$\rho>\rho\ICM$ are shown. In the shaded regions the gas is still
bound to the galactic potential. The size of the small rectangle in
the top left corner corresponds to $10\times 10$ grid cells.}
\label{fig:contours}%
\end{figure}

\begin{figure}
\resizebox{\hsize}{!}{\includegraphics{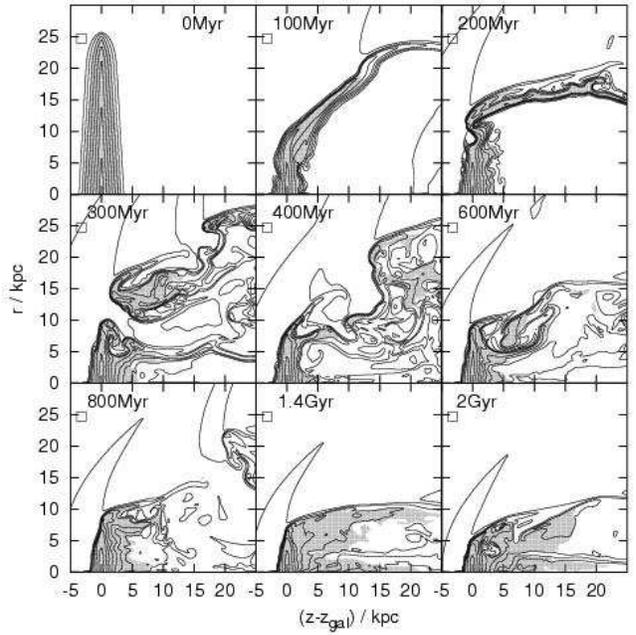}}
\caption{Like Fig.~\ref{fig:contours}, but for a wind with 
$\tilde p\Ram=1000\Rampresunit$, Mach 2.53. Again contours for
$\rho>\rho\ICM$ are shown. Here $\rho\ICM$ is lower than in
Fig.~\ref{fig:contours}, hence there are more contour lines in this
plot. As this is a supersonic wind, also a bow shock appears.}
\label{fig:contours_supersonic}%
\end{figure}

\begin{figure}
\resizebox{\hsize}{!}{\includegraphics{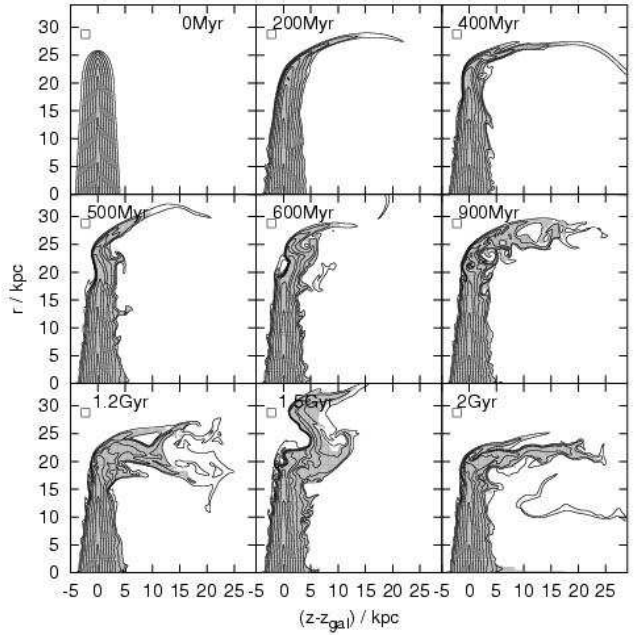}}
\caption{Like Fig.~\ref{fig:contours}, but for a weak wind with 
$\tilde p\Ram=10\Rampresunit$, Mach 0.8. Concerning the number of
contour lines, see the comment at Fig.~\ref{fig:contours_supersonic}.}
\label{fig:contours_weakwind}%
\end{figure}

%
To give an impression of the temporal evolution of the stripping
process in our simulations, we show the gas density distribution at
several time steps for a strong subsonic, a strong supersonic and a
weak subsonic wind in
Figs.~\ref{fig:contours},\ref{fig:contours_supersonic} and
\ref{fig:contours_weakwind}, respectively.

We find that the ICM-ISM stripping proceeds in three phases:
\begin{description}
\item[Instantaneous stripping:] First the outer gas disk, down to a 
     certain radius, is bent towards the downstream side of the
     galaxy. This phase can be seen in the first panel with $t>0\Myr$
     in Figs.~\ref{fig:contours} to \ref{fig:contours_weakwind}. The
     time scale on which the bending takes place is determined by the
     ram pressure. It reaches from a few $10\Myr$ for high ram
     pressures to about $200\Myr$ for low ram pressures. Also the
     radius down to which the disk is bent depends on the strength of
     the ram pressure. Reasonably, for higher ram pressures only a
     small inner part is retained, and for low ram pressures only the
     outer edge is bent, but the disk is affected even for very small
     ram pressures. In all cases a large part of the matter that is
     pushed out of its original position is still bound.
\item[Intermediate phase:] The first phase is followed by a quite 
     dynamical intermediate phase, during which the bent-out part of
     the disk breaks up. A part of the material that is pushed out is
     now stripped completely and a part falls back to the original
     disk region. This back-fall always happens in the lee of the
     galaxy, where the displaced gas is protected against the
     wind. This phase in which bound gas can be found behind the gas
     disk was also noticed by \citet{schulz01}. The duration of this
     phase depends on the ram pressure. For high ram pressures it is
     short (a few $100\Myr$), for low ram pressures this phase
     continues for the rest of the simulation (up to $2\Gyr$).
\item[Continuous stripping:] Actually the continuous stripping --the 
     peeling off of the outer disk layers of the upstream side of the
     gas disk by the Kelvin-Helmholtz (KH) instability -- works on the
     gas disk from the beginning on. However, the effect of this
     process is quite small, so that it is not distinct before the
     first two phases are finished. This status is only reached for
     stronger ram pressures, as for weaker ones the intermediate phase
     is long. Even though the effect of the continuous stripping is
     small, it slowly decreases the mass and the radius of the
     remaining gas disk.
\end{description}
Typical for the supersonic runs is a stagnant region in the wake of
the galaxy, where gas with low densities, that is still bound to the
galaxy, tends to linger (see final stages in
Fig.~\ref{fig:contours_supersonic}).

In all runs bound gas can be found at quite large distances (10 to
$15\,\Kpc$) behind the galaxy during the instantaneous stripping phase
and the intermediate dynamic phase. For lower ram pressures some gas
packages are still bound at distances of $20\,\Kpc$ behind the galaxy.

Another feature that appears in all simulations is the flittering of
the outer edge of the galaxy. The KH-instability pulls low density
tongues along the upstream side of the remaining gas disk from the
centre towards the edge. There these pieces of gas are exposed to the
full wind and blown away (see e.g.~last two panels in
Fig.~\ref{fig:contours}). This procedure repeats continuously and
leads to an oscillation in the radius of the gas disk $r\Disk(t)$.

\begin{figure}
\resizebox{\hsize}{!}{
\includegraphics[angle=-90]{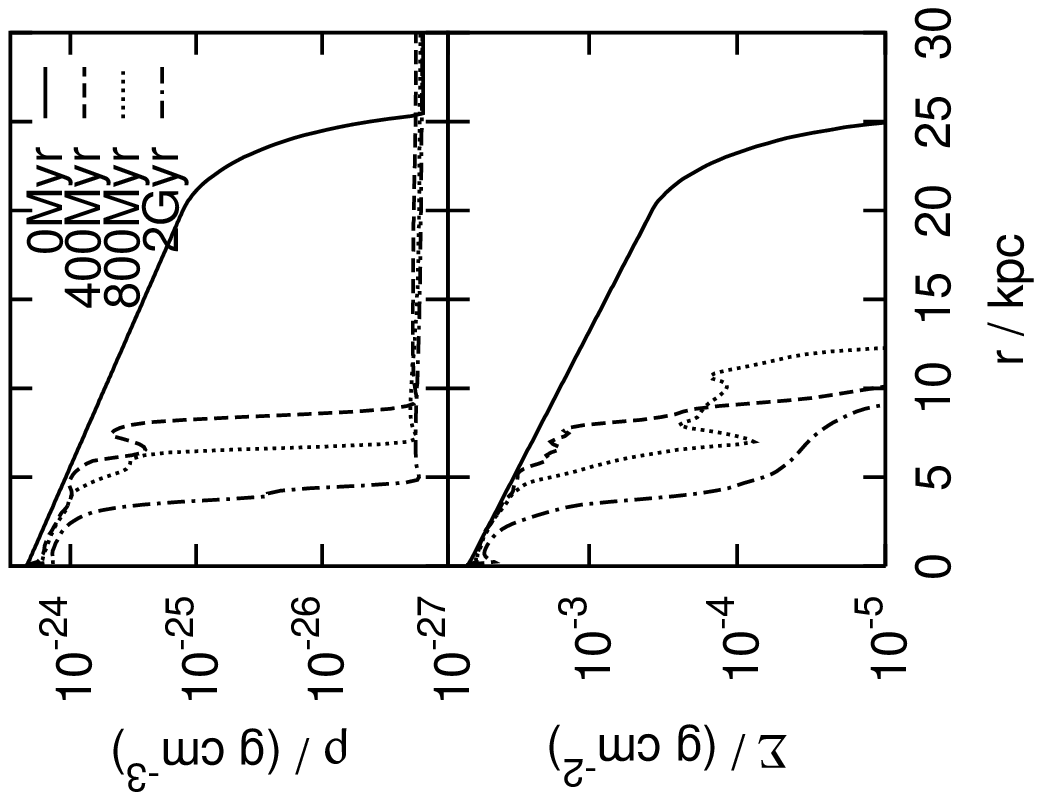}%
\includegraphics[angle=-90]{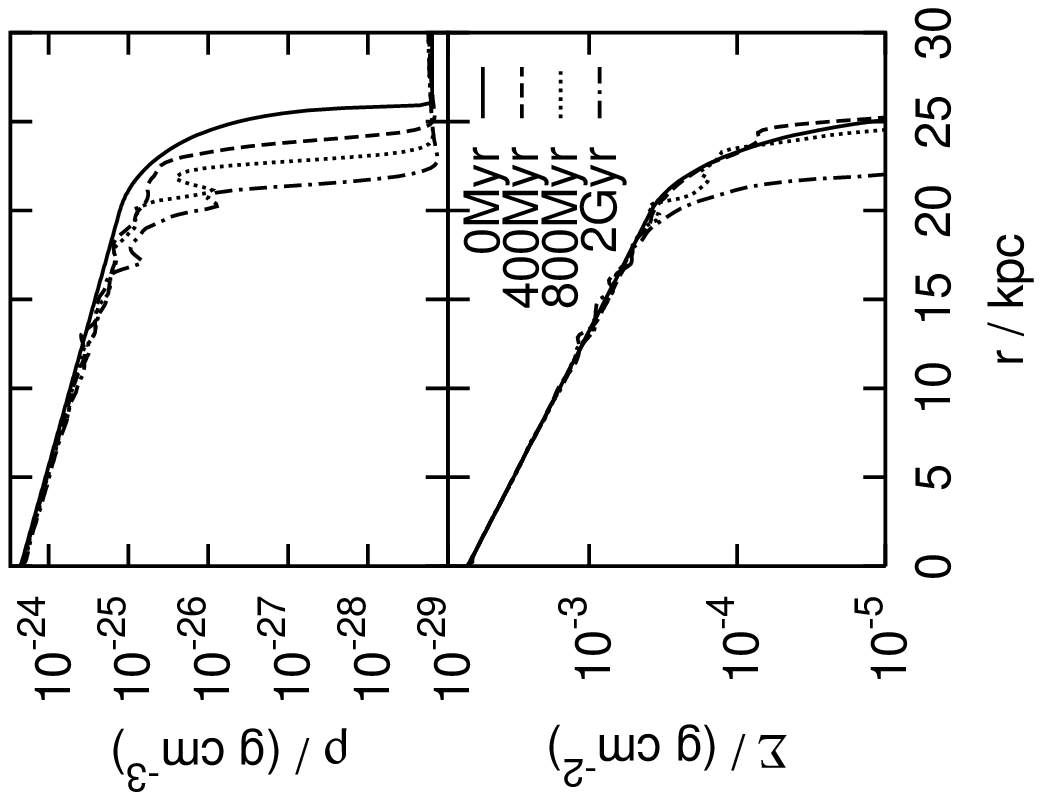}}
\caption{Evolution of density $\rho$ and surface density $\Sigma$ profiles along the galactic plane for to the runs shown in Figs.~\ref{fig:contours} (left) and \ref{fig:contours_weakwind} (right). For the calculation of $\Sigma$ we integrate over interval $(z\Gal-5\!\Kpc,z\Gal+5\!\Kpc)$ to avoid that stripped material at larger distances to the galaxy appears in $\Sigma$.}
\label{fig:profiles_evol}%
\end{figure}

The evolution of the density in the galactic plane and the surface
density profile are shown in Fig.~\ref{fig:profiles_evol} for a
stronger and for a weak subsonic wind.
As long as a substantial part of the gas disk remains, neither the
density profile nor the surface density in the inner part change
much. However, we observe a compression of the outer layers on the
upstream side of the gas disk (see
Sect.~\ref{sec:result_thickness}). For low ram pressures only the very
outer layers are compressed, for stronger winds the compression goes
deeper towards the galactic plane.
%
%
\subsection{Varying the wind parameters} \label{sec:result_wind}
\begin{figure}
\centering
\resizebox{0.9\hsize}{!}%
{\includegraphics[angle=-90]{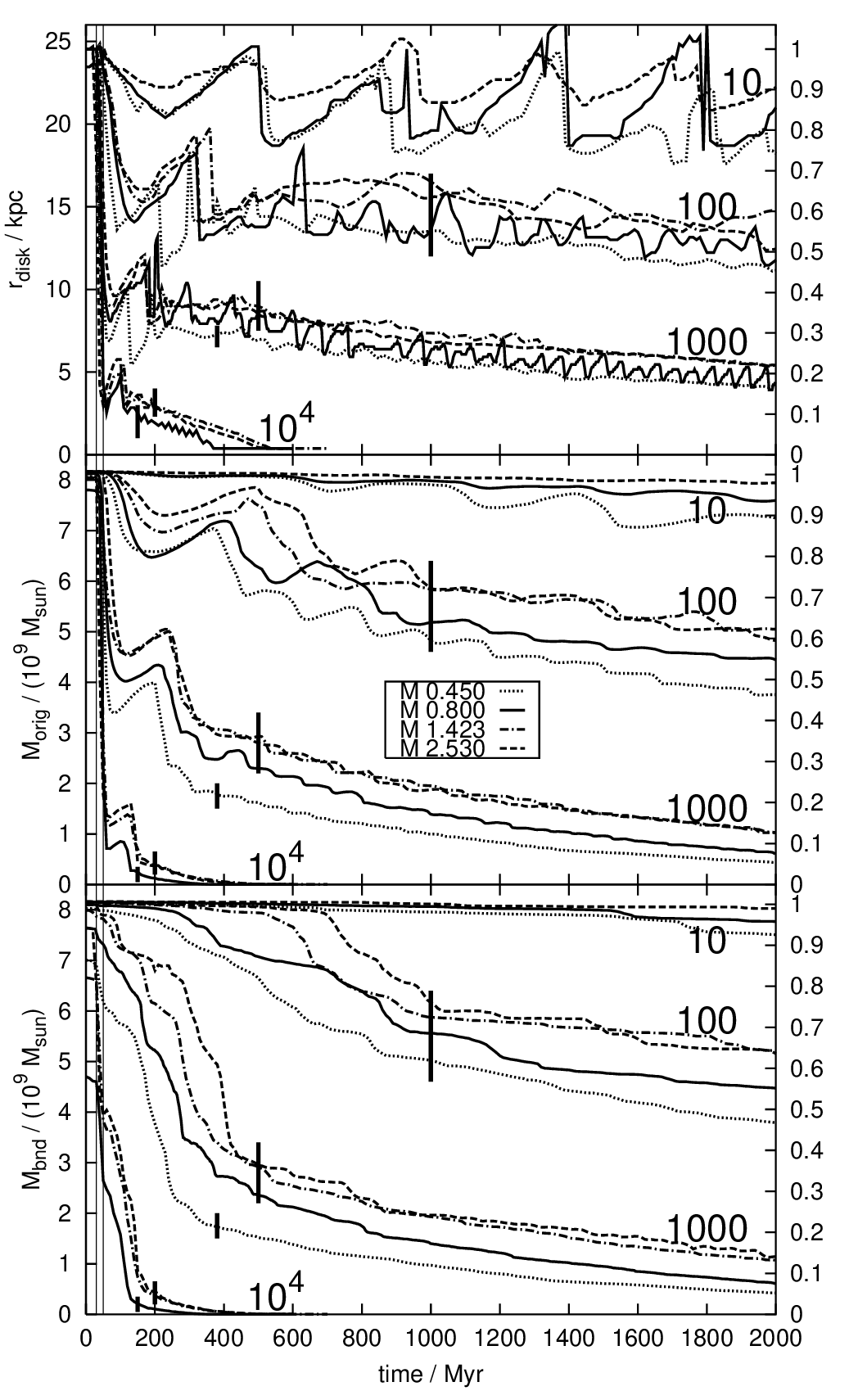}}
\caption{Gas disk radius $r\Disk(t)$ (top panel), mass $M\Origreg(t)$ in 
the original disk region (middle panel) and bound gas mass
$M\Bound(t)$ (bottom panel) of the massive galaxy in ICM winds with
$T\ICM{}_1$ for different ram pressures $\tilde p\Ram$ and Mach
numbers. Curves for the same $\tilde p\Ram$ group together. Each group
is labelled accordingly with its $\tilde p\Ram$ in units of
$\Rampresunit$. The Mach number of each curve is coded by the line
style, see key in the middle panel. The right axes are labelled in
units of the initial radius or mass.\newline 
In the top panel all
curves except the ones for Mach 0.8 are shown in the smoothed version
(see text) for clarity. The frequency and amplitude in $r\Disk(t)$ are
comparable for all runs with the same $\tilde p\Ram$.\newline 
The vertical bars show where we consider the intermediate phase to be
finished. The two thin vertical lines mark the moment when the flow
reaches the galaxy ($t_0=40\Myr$) and $t_0+t\Switch=60\Myr$.}
\label{fig:comp-masses-radii}%
\end{figure}

\begin{figure}
\centering
\resizebox{\hsize}{!}{\includegraphics[angle=-90]{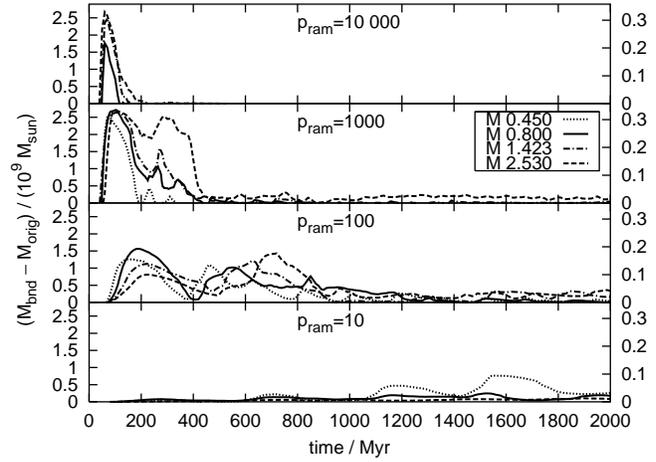}}
\caption{Mass of bound gas outside the original disk region as a function 
of time; for the massive galaxy in different ICM winds
($T\ICM{}$). There is one labelled panel for each ram pressure. Inside
the panels the Mach numbers of the wind is coded by the line style
(see key). The right axes are labelled in units of the initial radius
or mass.}
\label{fig:comp-bound}%
\end{figure}

\begin{figure}
\centering
\resizebox{\hsize}{!}{\includegraphics[angle=-90]{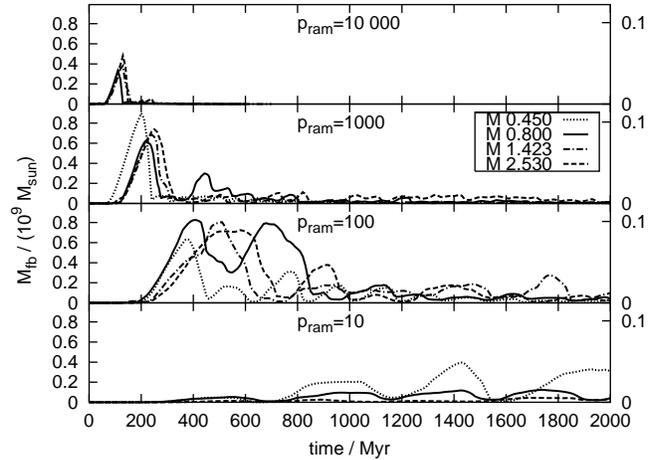}}
\caption{Mass of fallen back gas $M\Fallback(t)$(= galactic gas that has 
left the original disk region and fell back to that region) as
function of time; for the massive galaxy in different ICM winds
($T\ICM{}_1$). There is one labelled panel for each ram
pressure. Inside the panels the Mach numbers of the wind is coded by
the line style (see key). The right axes are labelled in units of the
initial radius or mass.}
\label{fig:comp-fallback}%
\end{figure}

%
In this Sect. we show how the results for the gas disk radius
$r\Disk(t)$, the gas mass inside the original disk region
$M\Origreg(t)$ and the bound gas mass $M\Bound(t)$ depend on the wind
strength. We also show in separate plots (in order to maintain
clarity) how much bound gas can be found outside the original disk
region (that is $M\Bound(t)-M\Origreg(t)$) and how much fallen back
gas is contained inside the original disk region (this is
$M\Fallback(t)$, see Sect.~\ref{sec:definitions}). We scan the wind
parameter space for three different cases:
\begin{itemize}
\item[(a)] The massive galaxy in different ICM winds with ICM temperature 
     $T\ICM{}_1$ (corresponding to an ICM sound speed of
     $1000\Kms$). 
\item[(b)] The massive galaxy in different ICM winds with ICM temperature 
     $T\ICM{}_2$ (corresponding to an ICM sound speed of
     $562\Kms$). 
\item[(c)] The medium galaxy in different ICM winds with ICM temperature 
     $T\ICM{}_1$ (corresponding to an ICM sound speed of
     $1000\Kms$). 
\end{itemize}
As a representative case, the results of case (a) are shown in
Figs.~\ref{fig:comp-masses-radii}, \ref{fig:comp-bound} and
\ref{fig:comp-fallback}. Corresponding plots for cases (a) and (b) are 
very similar.
%
%
%
%
For clarification we apply a time-averaging to most $r\Disk(t)$
curves, smoothing the oscillations that are due to the flittering edge
of the galaxy (see end of Sect.~\ref{sec:result_general}). We average
$r\Disk(t)$ over $100\Myr$ for the cases with $\tilde p\Ram\ge
1000\Rampresunit$ ($\tilde p\Ram\ge 100\Rampresunit$ for the medium
galaxy) and over $200\Myr$ for the lower ram pressure cases. The
smoothing is applied neither to the initial $250\Myr$ ($500\Myr$) for
the stronger ram pressures (for the weaker ram pressures), nor to the
curves for $\tilde p\Ram= 10\Rampresunit$. Also in each group of
constant ram pressure we leave one curve oscillating. The amplitude
and the frequency of the oscillations are similar within a group.

The first thing to notice is that the mass and radius of the remaining
gas disk depend mainly on the ram pressure with higher ram pressures
resulting in smaller disks. For the strongest ram pressures even the
complete gas disk is stripped. At second glance one can see that for a
given ram pressure the gas loss is stronger for subsonic winds than
for supersonic ones.

The phases of the stripping process as they were explained in
Sect.~\ref{sec:result_general} are clearly visible in the $r\Disk(t)$,
$M\Origreg(t)$ and $M\Bound(t)$ curves. The bending of the outer gas
disk during the instantaneous stripping phase results in a first
strong decrease of $r\Disk(t)$ and $M\Origreg(t)$. We define the
duration of this phase $\tau_{\mathrm{push}}$ to last from the moment
the flow reaches the galaxy to the first local minimum in
$r\Disk(t)$. Fig.~\ref{fig:timescales} shows the dependence of
$\tau_{\mathrm{push}}$ on ram pressure, this point is discussed in
Sect.~\ref{sec:result_comp_an-num_push}. After the instantaneous
stripping phase, the intermediate dynamic phase follows with the
falling back of some gas. This can be seen in the slight increase of
$M\Origreg(t)$ just after the initial decrease, and of course in the
plot showing $M\Fallback(t)$ directly
(Fig.~\ref{fig:comp-fallback}). Although the gas disk always reacts to
the wind in the initial instantaneous stripping phase, only for
stronger winds this also results in unbinding a substantial amount of
gas from the galactic potential. The duration of the intermediate
phase, this is until most of the bound gas outside the disk region has
vanished, takes much longer for weak winds than for strong ones. For
ram pressures of $10\Rampresunit$ this phase is not finished at the
end of the simulations. For all but the very weakest ram pressure,
during the first few $100\Myr$ between 20 and 40\% of the original gas
disk mass linger outside the original disk region but are still
bound. The amount of gas that falls back to the original disk region
is of the order of 5 to 10\% of the original gas disk mass.

For the weakest ram pressure ($\tilde p\Ram=10\Rampresunit$) the
galaxy loses only a few percent of its gas, although also in this case
the radius of the gas disk is truncated.

\subsection{Influence of the wind initialisation} \label{sec:result_tswitch}
\begin{figure*}
\resizebox{\hsize}{!}{
\includegraphics[height=0.5\textwidth,angle=-90]{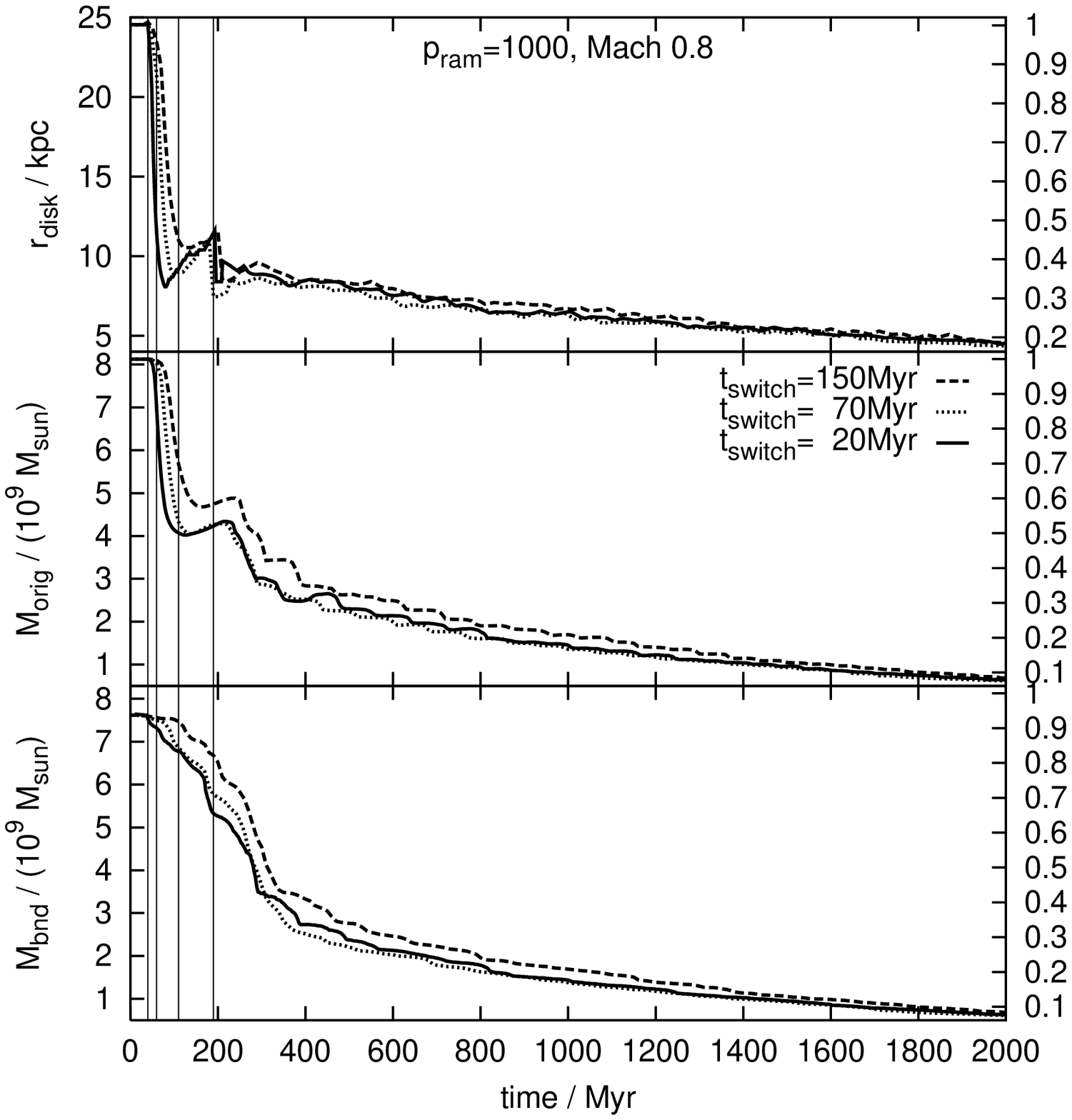}
\includegraphics[height=0.5\textwidth,angle=-90]{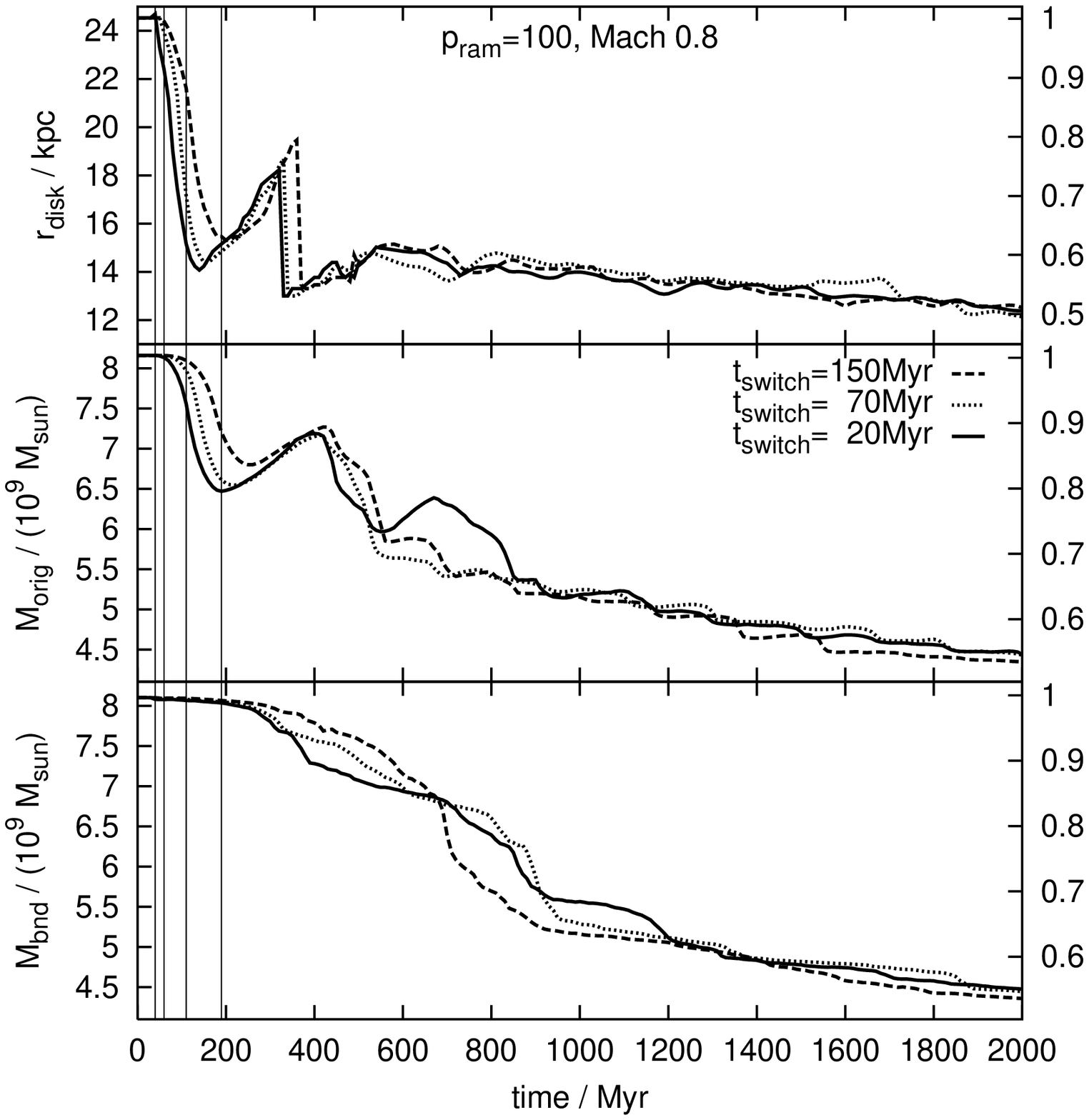}}
\caption{Comparison of stripping radius $r\Disk(t)$, mass in original disk 
region $M\Origreg(t)$ and bound mass $M\Bound(t)$ for different
$t\Switch$, for two representative winds. We marked the moment where
a sound wave starting at the inflow boundary reached galaxy ($t_0$),
and $t_0+t\Switch$ for the three different $t\Switch$ by the four thin
vertical lines. For further explanations see
Fig.~\ref{fig:comp-masses-radii}.}
\label{fig:result_tswitch}%
\end{figure*}

%
We test how our results are influenced by the initialisation of the
simulations (see Sect.~\ref{sec:sim_tswitch}). In addition to the
standard case $t\Switch=20\Myr$ we rerun a few representative cases
with $t\Switch=70\Myr$ and $t\Switch=150\Myr$. The resulting mass and
radius of the gas disk for two of these winds are shown in Fig.~\ref{fig:result_tswitch}.
Again we smooth the curves for $r\Disk(t)$ (see
Sect.~\ref{sec:result_wind}). Different $t\Switch$ result in expected
systematic differences during the initial phase. The decrease in mass
and radius due to the instantaneous stripping is fastest for the
shortest $t\Switch$. After that $r\Disk$, $M\Origreg$ and $M\Bound$
agree well despite the different $t\Switch$. The cases with the
longest $t\Switch$ tend to retain a little more mass during the
intermediate phase. For weaker $\tilde p\Ram$ the agreement between
cases with different $t\Switch$ is not as tight as for strong $\tilde
p\Ram$, which is due to the longer dynamic (chaotic) intermediate
phase for weaker $\tilde p\Ram$. The qualitative behaviour of the
amount of bound mass outside the disk region and of the fallen back
mass is independent of $t\Switch$.
%
%
\subsection{Influence of the vertical structure of the gas disk}
\label{sec:result_thickness}
\begin{figure}
\resizebox{\hsize}{!}{\includegraphics{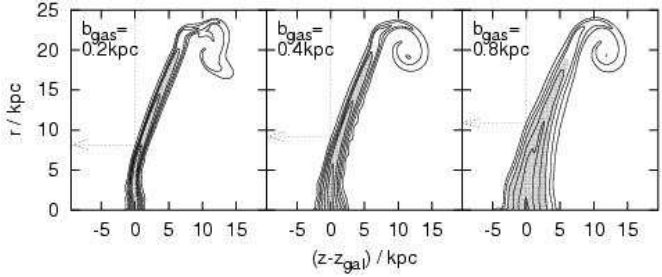}}
\caption{Density contours at 70Myr for three different gas disk 
thicknesses ($b\Gas=0.2,0.4,0.8\Kpc$). Gas in the shaded region is
bound to the galactic potential. The wind was the same in all cases
($\tilde p\Ram=1000\Rampresunit$, $T\ICM{}_1$, Mach 0.8). The method
of radius measurement is illustrated (see
Sect.~\ref{sec:definitions}).}
\label{fig:denscont_thickness}%
\end{figure}

\begin{figure*}
\resizebox{\hsize}{!}{
\includegraphics[height=0.5\textwidth,angle=-90]{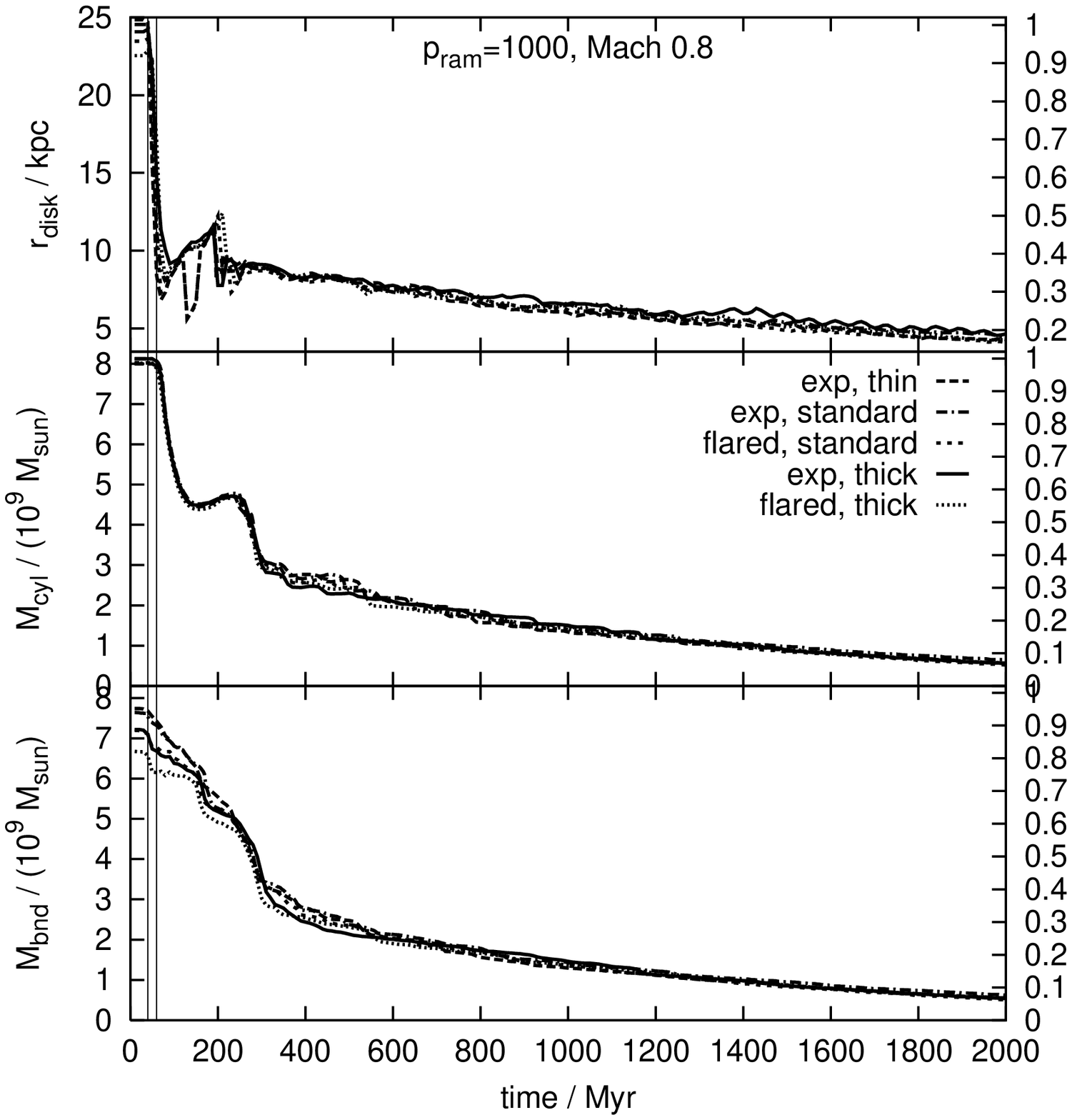}
\includegraphics[height=0.5\textwidth,angle=-90]{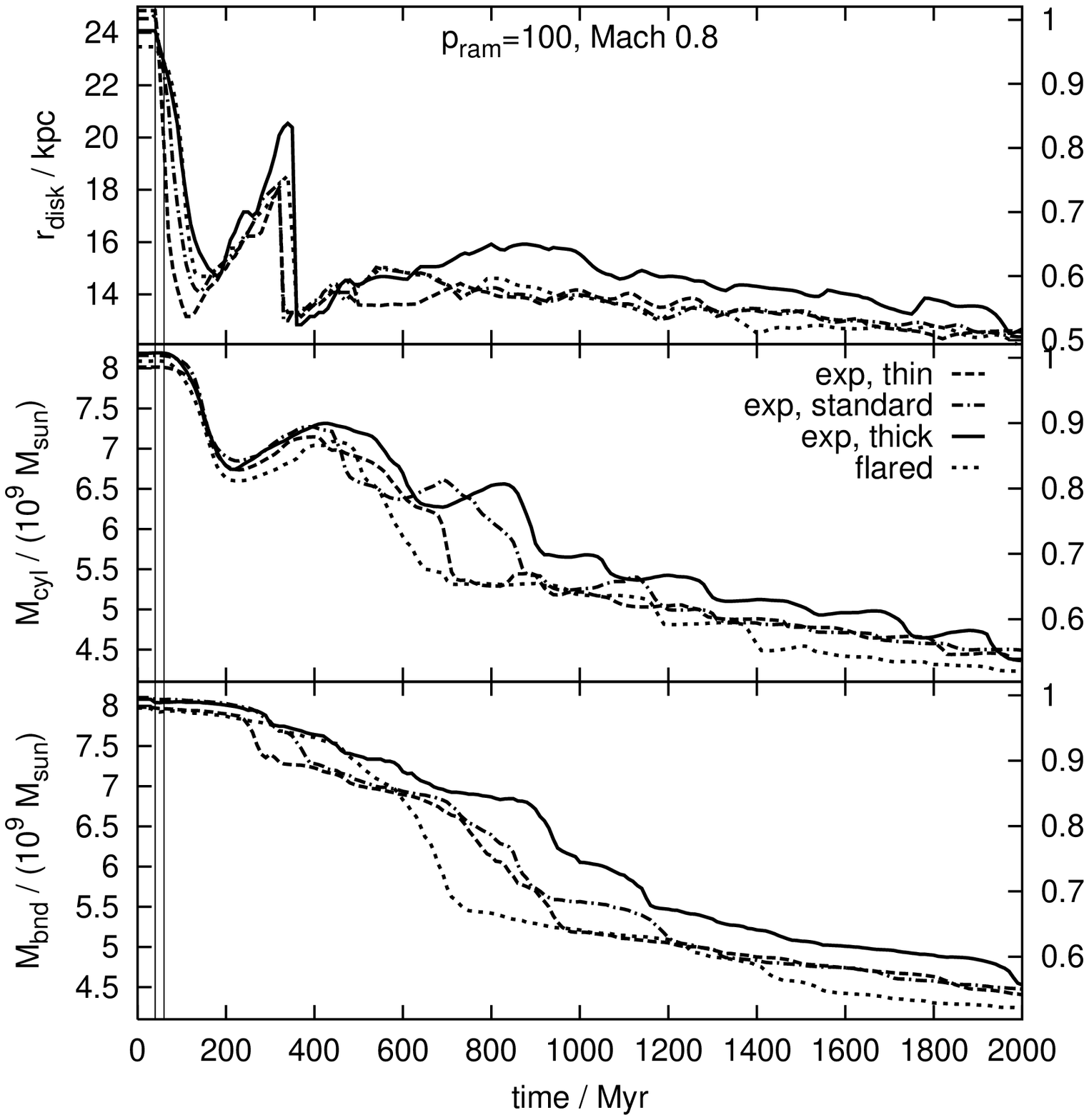}}
\caption{Comparison of stripping radius $r\Disk(t)$, mass in original 
disk region $M\Origreg(t)$ and bound mass $M\Bound(t)$ for different
vertical structures, for two representative winds. For further
explanations see Fig.~\ref{fig:comp-masses-radii}. All radius curves
are smoothed as explained in Sect.~\ref{sec:result_wind}.}
\label{fig:result_thickness}%
\end{figure*}

In order to study the influence of the vertical structure of the gas
disk we perform some simulation runs with exponential and flared disks
of varying scale heights in representative wind cases (see
Table~\ref{tab:runs_thickness}). Fig.~\ref{fig:denscont_thickness}
shows snapshots at the same moment during the instantaneous stripping
phase for three exponential disks with different scale heights.
Despite their different thickness, all three cases look very similar.
The resulting radius and mass curves are
compared in Fig.~\ref{fig:result_thickness} for two of the winds.
It does not make sense to use $M\Origreg$ for this comparison, as for
gas disks with different thicknesses and shapes also the original disk
regions differ strongly. Hence we use $M\Cylreg$ here, the mass inside
the cylinder with $r<26\Kpc$ and $|z-z\Gal|<5\Kpc$. The different
starting points of the curves for the bound mass are due to the
different disk thicknesses and external pressures (see
Sect.~\ref{sec:ini_gasdisk} and Fig.~\ref{fig:comp_pres}); for thicker
disks the outer layers become unbound.  The radius curves are again
smoothed as explained in Sect.~\ref{sec:result_wind}.

The $r\Disk$ as well as $M\Cylreg$ and $M\Bound$ are nearly
independent of the vertical structure of the gas disk. For the higher
ram pressure of $1000\Rampresunit$ the results for the different gas
disks agree excellently. For the lower ram pressure
($100\Rampresunit$) the scatter is a bit larger as the galaxies stay
longer in the dynamic intermediate phase. The difference in the radius
curves during the initial instantaneous stripping phase are due to the
method of measurement, as is demonstrated in
Fig.~\ref{fig:denscont_thickness}.  We conclude from this test that
the success of RPS is independent of the vertical structure of the gas
disk.

\begin{figure}
\resizebox{\hsize}{!}{\includegraphics[angle=-90]{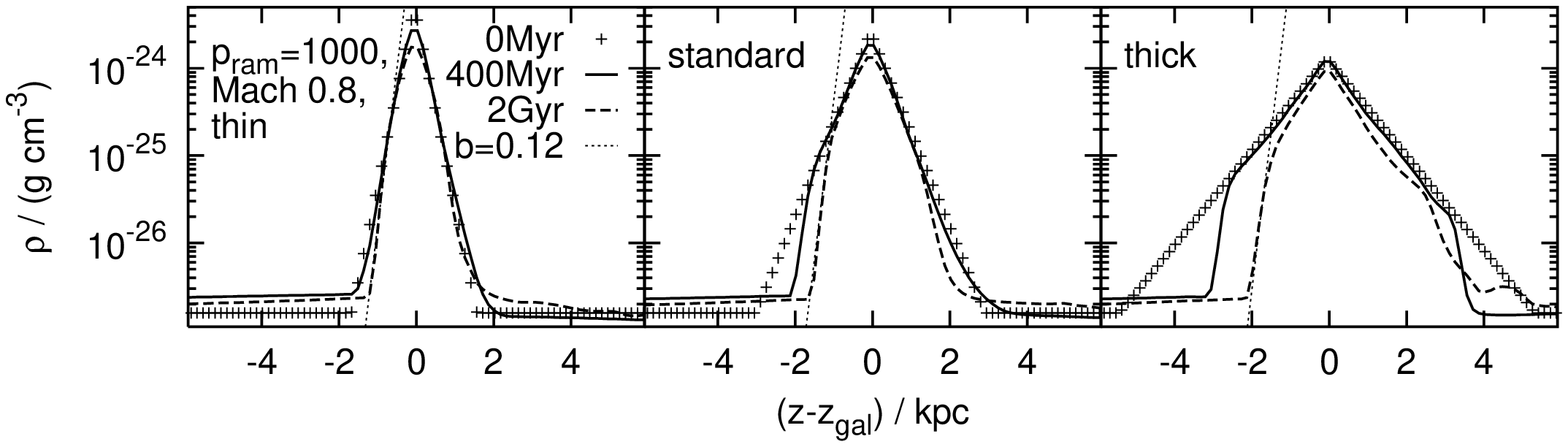}}
\resizebox{\hsize}{!}{\includegraphics[angle=-90]{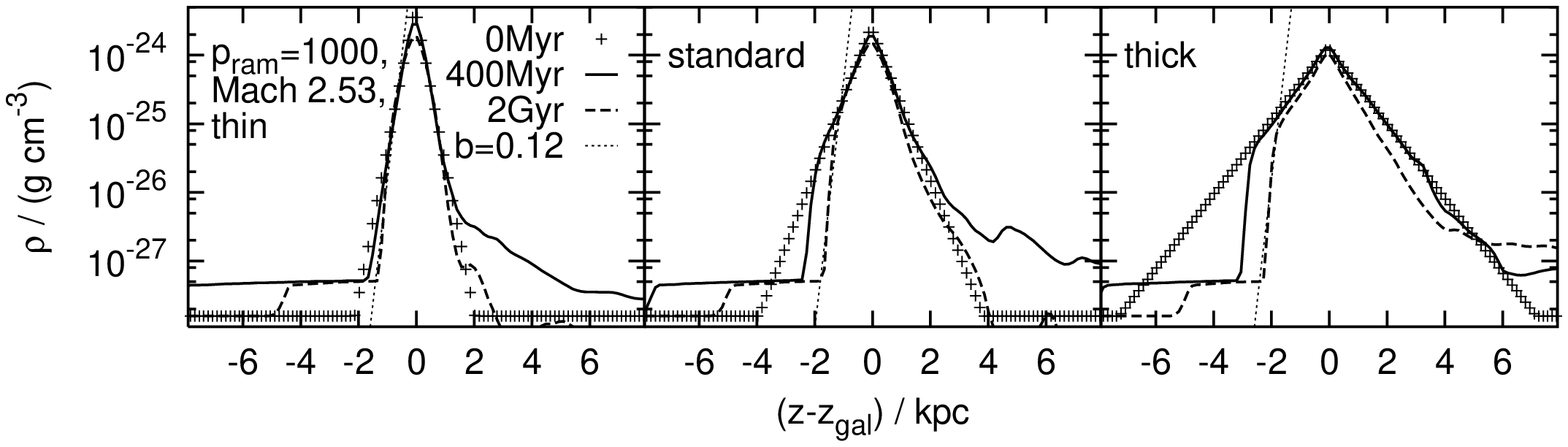}}
\resizebox{\hsize}{!}{\includegraphics[angle=-90]{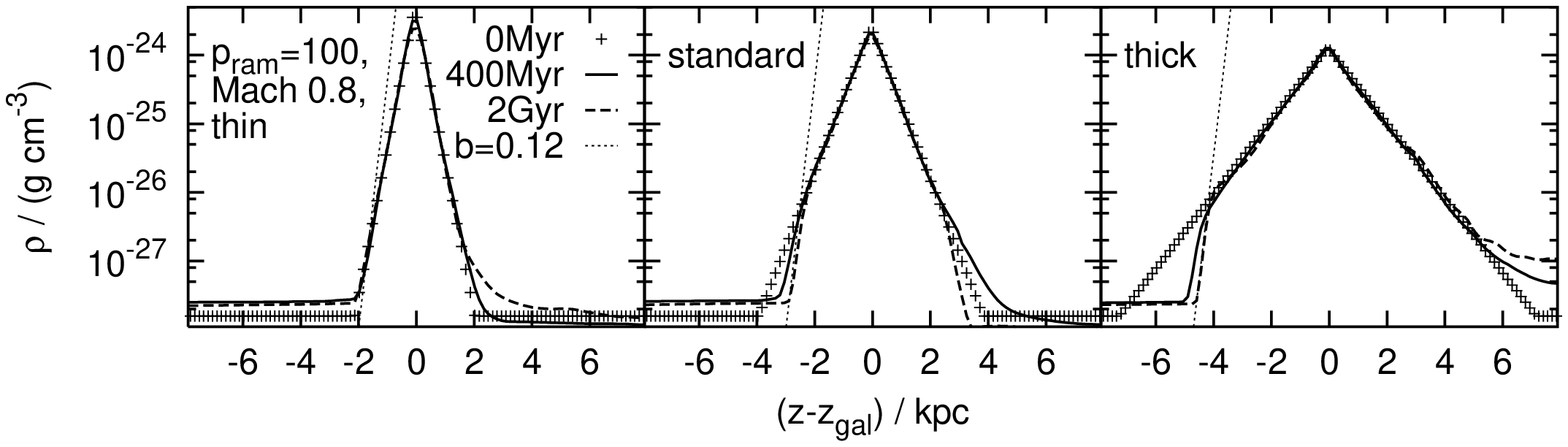}}
\resizebox{\hsize}{!}{\includegraphics[angle=-90]{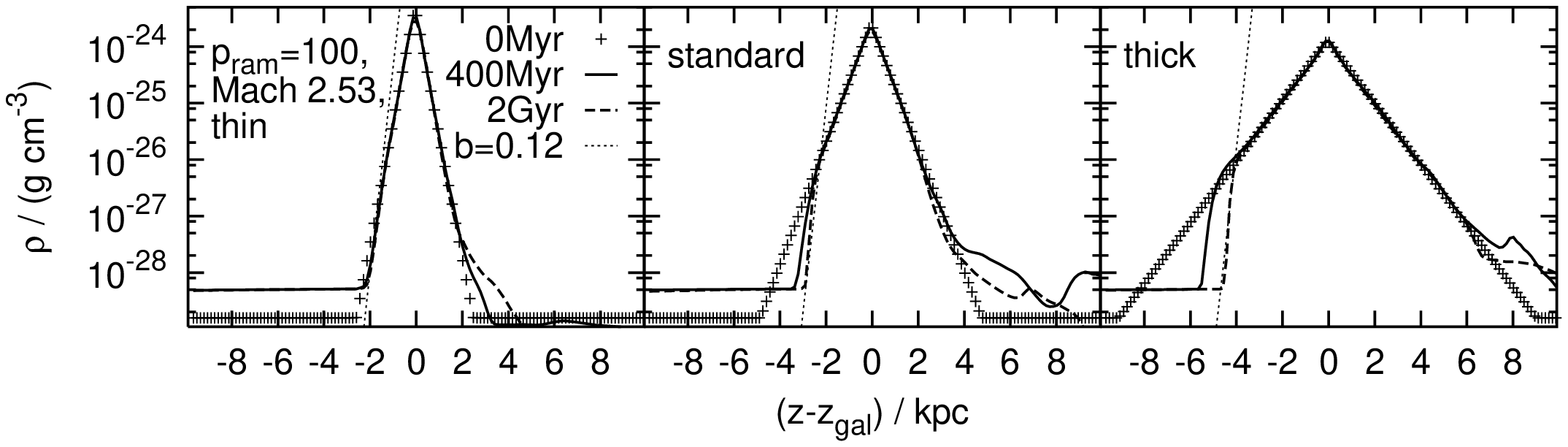}}
\resizebox{\hsize}{!}{\includegraphics[angle=-90]{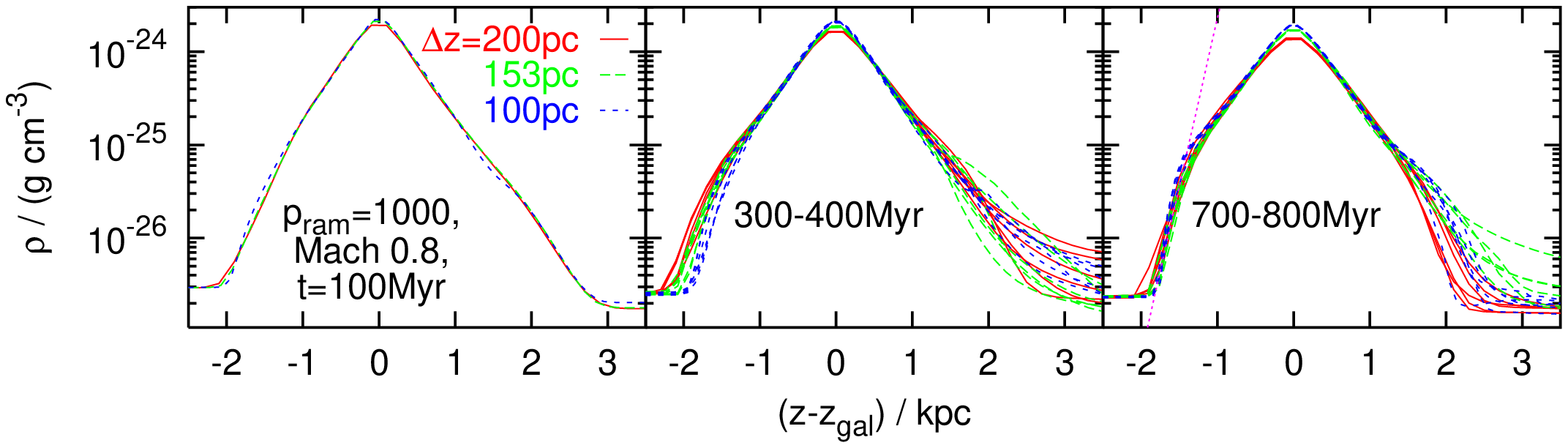}}
\caption{Evolution of vertical profiles at $r=0.23\Kpc$ for different 
winds and gas disk thickness. We used an exponential disk for all
cases, the scale heights were $b\Gas=0.2\Kpc$ (thin), $b\Gas=0.4\Kpc$
(standard) and $b\Gas=0.8\Kpc$ (thick). We fitted an exponential
function to the steepened part of the profile, the scale height for
this fit was approximately $0.12\Kpc$ for all cases. In the bottom
line the variation of the vertical profiles with resolution for three
different time steps/time intervals is demonstrated (colour version in
online paper).}
\label{fig:vprofiles}%
\end{figure}

We compare the compression of the upstream side of the gas disk for
different scale heights and winds by showing the evolution of the
vertical density profile through the gas disk at $r=0.23\Kpc$ in
Fig.~\ref{fig:vprofiles}.
The compression of the upstream side of the gas disk results in a
steepening of the density profile of the outer layers. In the
compressed region the density profile can be fitted with an
exponential function of a smaller scale height than the original
$b\Gas$. The scale height in the compressed layers is about $0.12\Kpc$
independently of the wind and the original $b\Gas$, but for stronger
ram pressures the compressed layer reaches deeper towards the galactic
plane. The dependence of the scale height of the compressed layer on
resolution is demonstrated in the bottom line in
Fig.~\ref{fig:vprofiles}, it is only a weak dependence. This plot
shows vertical profiles at $t=100\Myr$ and in the time intervals
$300$-$400\Myr$ and $700$-$800\Myr$. We chose to show time intervals
to allow averaging over the intrinsic oscillations in the profiles. In
any way, if the scale height of the compressed layer was set by the
resolution, it should vary from thick to thin disks.  For the medium
galaxy the profile of the outer layers is steepened to a scale height
of about 0.09kpc.
%
%
\subsection{Cross-comparisons between runs with different $T\ICM$}
\label{sec:result_comparison_TICM}
\begin{figure*}
\resizebox{\hsize}{!}{
\includegraphics[height=\textwidth,angle=-90]{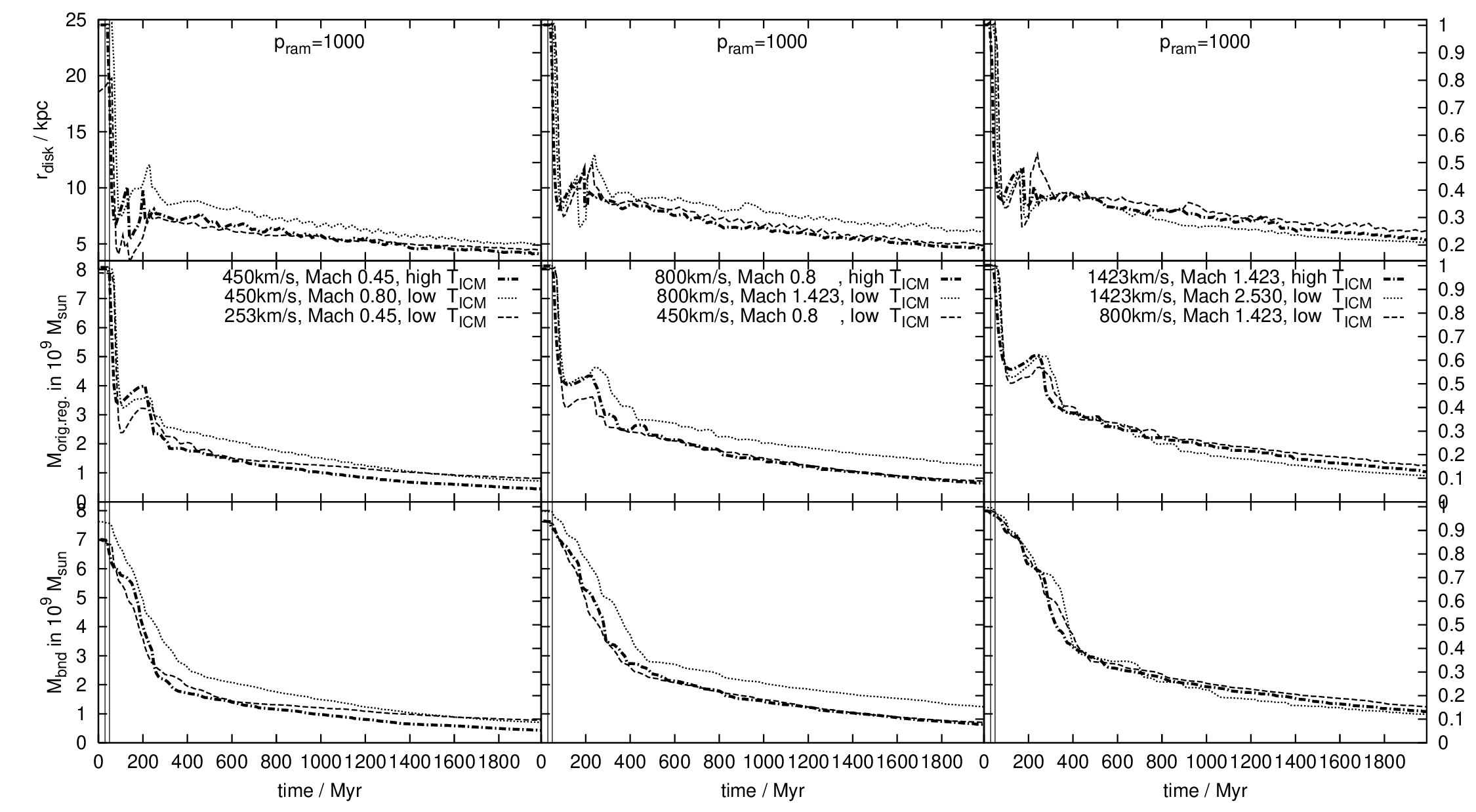}}
\resizebox{\hsize}{!}{
\includegraphics[height=\textwidth,angle=-90]{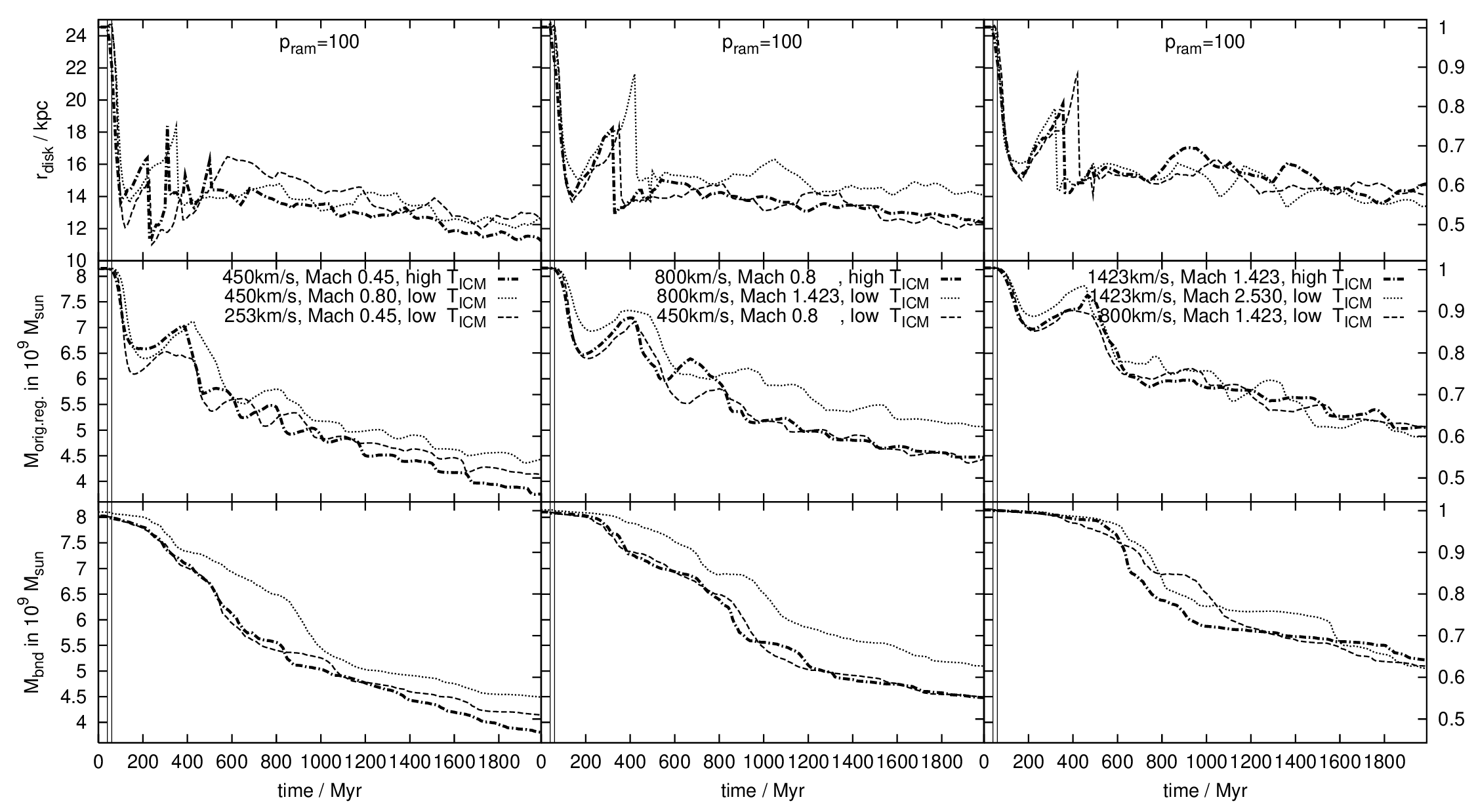}}
\caption{Comparison of stripping radius $r\Disk(t)$, mass in original 
disk region $M\Origreg(t)$ and bound mass $M\Bound(t)$ for selected
runs with $\tilde p\Ram = 1000\Rampresunit$ (upper panel) and $\tilde
p\Ram100=\Rampresunit$ (lower panel). For further explanations see
Fig.~\ref{fig:comp-masses-radii}. See also
Sect.~\ref{sec:result_comparison_TICM}.}
\label{fig:result_Machnumber}%
\end{figure*}

The results from Sect.~\ref{sec:result_wind} suggest that the effect
of the stripping does not depend on the ram pressure alone but also
slightly on the Mach number. To investigate this point further we
compare runs with different ICM temperatures. E.g.~for $T\ICM{}_1$ the
velocity of $800\Kms$ is equivalent to Mach number 0.8, whereas for
$T\ICM{}_2$ the same velocity corresponds to Mach number 1.423. The
velocity that belongs to Mach number 0.8 for $T\ICM{}_2$ is
$450\Kms$. If indeed the Mach number is the determining parameter,
then for this example the result of the run with
$(T\ICM{}_1,800\Kms,\textrm{Mach number}\,0.8)$ should be the same as
the result of the case $(T\ICM{}_2,450\Kms,\textrm{Mach number}\,0.8)$
rather than $(T\ICM{}_2,800\Kms,\textrm{Mach number}\,1.423)$. We show
this comparison for $r\Disk(t)$, $M\Origreg(t)$ and $M\Bound(t)$ in
Fig.~\ref{fig:result_Machnumber} for some cases.
In the middle panel, where the velocity is subsonic for $T\ICM{}_1$
and supersonic for $T\ICM{}_2$, indeed runs with the same Mach number
rather than with the same velocity correspond to each other. The same
seems to hold for the left panels, in the pure subsonic range,
although not as clear as in the transition from the subsonic to the
supersonic regime. In the supersonic range (right panels) no
systematic difference is seen. As qualitative issues we conclude that
(i) for a given wind density and velocity, RPS is slightly more
effective if this wind is subsonic (compared to supersonic winds); and
(ii) that in the subsonic regime winds with smaller Mach numbers are a
bit more effective than winds with higher Mach numbers. Nevertheless,
the Mach number is only a secondary parameter, the main parameter is
the ram pressure.

Moreover, we do not want to over-interpret the differences between the
subsonic and supersonic cases, as we used a single-fluid
description. In contrast, the ICM is highly ionised and actually
consists of a proton and an electron fluid. With respect to the proton
fluid, the galaxy's velocities are transonic, but with respect to the
electron fluid they still move subsonically. \citet{portnoy93} studied
the difference between the single-fluid and a two-fluid description
for a spherical galaxy (with gas replenishment) and found no
difference for the amount of gas remaining inside the galaxy.
%
%
\subsection{Comparison of analytical estimate and numerical result}
\label{sec:result_comparison_analytic-numeric}
%
\subsubsection{Instantaneous stripping} \label{sec:result_comp_an-num_push}
We have found that during the instantaneous stripping phase the outer
part of the gas disk is displaced, but not immediately unbound. It
takes a while until the displaced gas is also unbound. As the
analytical estimate in Eq.~\ref{eq:gunngott} just compares the forces
working on the gas disk, it tells how much gas will be unbound, but it
cannot quantify how long this process may take. As the analytical
estimate states how much gas will be unbound from the galaxy, we also
need to extract this quantity from the simulations. Hence we measure
the radius and the mass of the remaining gas disk at the end of the
dynamic intermediate phase, i.e.~when no substantial amount of bound
gas outside the disk region is left. We have marked this moment in
Fig.~\ref{fig:comp-masses-radii} by vertical bars. For the lowest ram
pressure ($\tilde p\Ram=10\Rampresunit$) the intermediate phase is not
completed during the simulation run time, here we use the final values
for the comparison.

\begin{figure}
\centering\resizebox{0.7\hsize}{!}{\includegraphics[angle=-90]{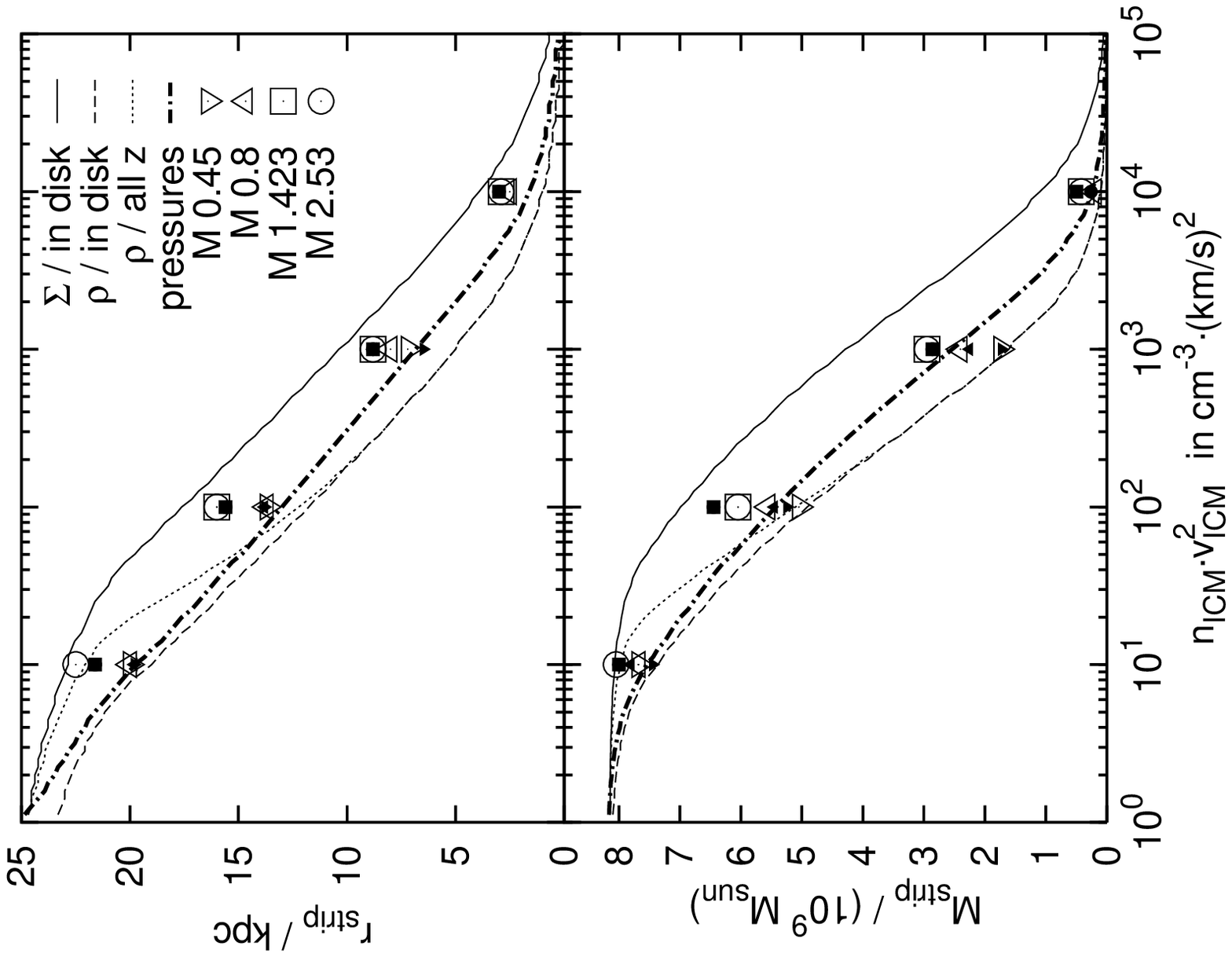}}
\caption{Comparison of analytical estimate and numerical result for the 
stripping radius and the remaining gas mass as a function of ram
pressure, for the massive galaxy. We show four versions of the
analytical estimate, see Sect.~\ref{sec:result_comp_an-num_push}. The
Mach numbers for the numerical result are coded with different symbol
shapes. Open symbols are for cases with $T\ICM{}_1$, solid symbols of
the same shape for cases with $T\ICM{}_2$.}
\label{fig:comp_an_num}%
\end{figure}

\begin{figure}
\centering\resizebox{0.7\hsize}{!}{\includegraphics[angle=-90]{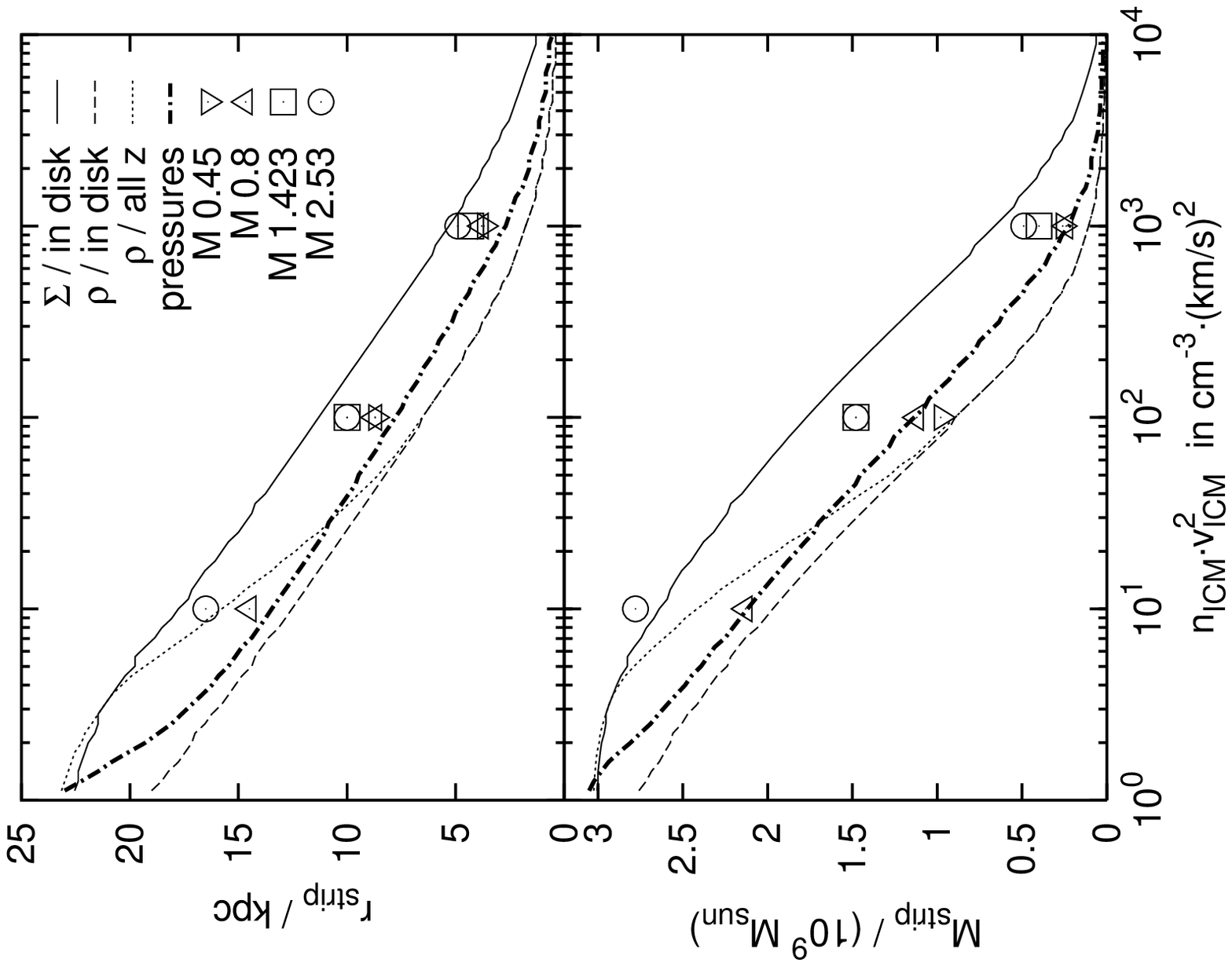}}
\caption{Same as Fig.~\ref{fig:comp_an_num}, but for the medium galaxy.}
\label{fig:comp_an_num_medgal}%
\end{figure}

As we pointed out in Sect.~\ref{sec:analytic_instantaneous}, there are
several versions how the analytical estimate can be done, therefore we
discuss the following recipes:
\begin{itemize}
\item[(a)] Use Eq.~\ref{eq:fgrav} to compute the restoring force while 
     using the maximal potential gradient inside the original disk
     region (marked by ``$\Sigma/$in disk'' in
     Figs.~\ref{fig:comp_an_num} and \ref{fig:comp_an_num_medgal}).
\item[(b)] Use Eq.~\ref{eq:fgrav2} with $\Delta z=150\PC$ and use the 
     maximal potential gradient inside the original disk region
     (marked by ``$\rho/$in disk'' in Figs.~\ref{fig:comp_an_num} and
     \ref{fig:comp_an_num_medgal}).
\item[(c)] Use Eq.~\ref{eq:fgrav2} with $\Delta z=150\PC$ and use the 
     maximal potential gradient found for any $z$ (marked by ``$\rho/$
     all $z$'' in Figs.~\ref{fig:comp_an_num} and
     \ref{fig:comp_an_num_medgal}).
\item[(d)] Compare the thermal pressure in the galactic plane and the 
     ram pressure (Eq.~\ref{eq:estimate_comp_pres}, marked by
     ``pressures'' in Figs.~\ref{fig:comp_an_num} and
     \ref{fig:comp_an_num_medgal})
\end{itemize}
Version (c) is between (a) and (b); for higher ram pressures it is
indistinguishable from version (b), as there smaller radii are
concerned where the steepest potential gradient is found inside the
disk region. Also the result from the pressure comparison (d) has a
very similar shape to (a) and (b) and is between those two. As already
mentioned in Sect.~\ref{sec:analytic_instantaneous}, the estimates
according to versions (a) and (b) are identical despite a shift in
horizontal direction for exponential disks. These two versions
represent the two extremes in the uncertainty how the gravitational
restoring force $f\Grav$ shall be computed -- either simplifying to a
thin disk (a) or computing $f\Grav$ for a single cell only (b). This
uncertainty can be parametrised by a factor of $\alpha$ in the
calculation of $f\Grav$, modifying Eq.~\ref{eq:fgrav} to
\be
|f\Grav(z,r)| = \alpha \left|\frac{\partial\Phi}{\partial z} (z,r) \Sigma\Gas(r)\right|. \label{eq:fgrav3}
\ee
Choosing $\alpha=1$ corresponds to version (a), choosing $\alpha=0.16$
recovers version (b), but $\alpha$ can be adjusted to fit the
numerical results.

Before we do any fitting, we show the pure analytical estimates
(versions (a) to (d)) and the numerical results in
Fig.~\ref{fig:comp_an_num} for the massive galaxy and in
Fig.~\ref{fig:comp_an_num_medgal} for the medium galaxy.
%
%
The numerical results are indeed in the range suggested by the
analytical estimates.

\begin{figure}
\centering\resizebox{0.7\hsize}{!}{\includegraphics[angle=-90]{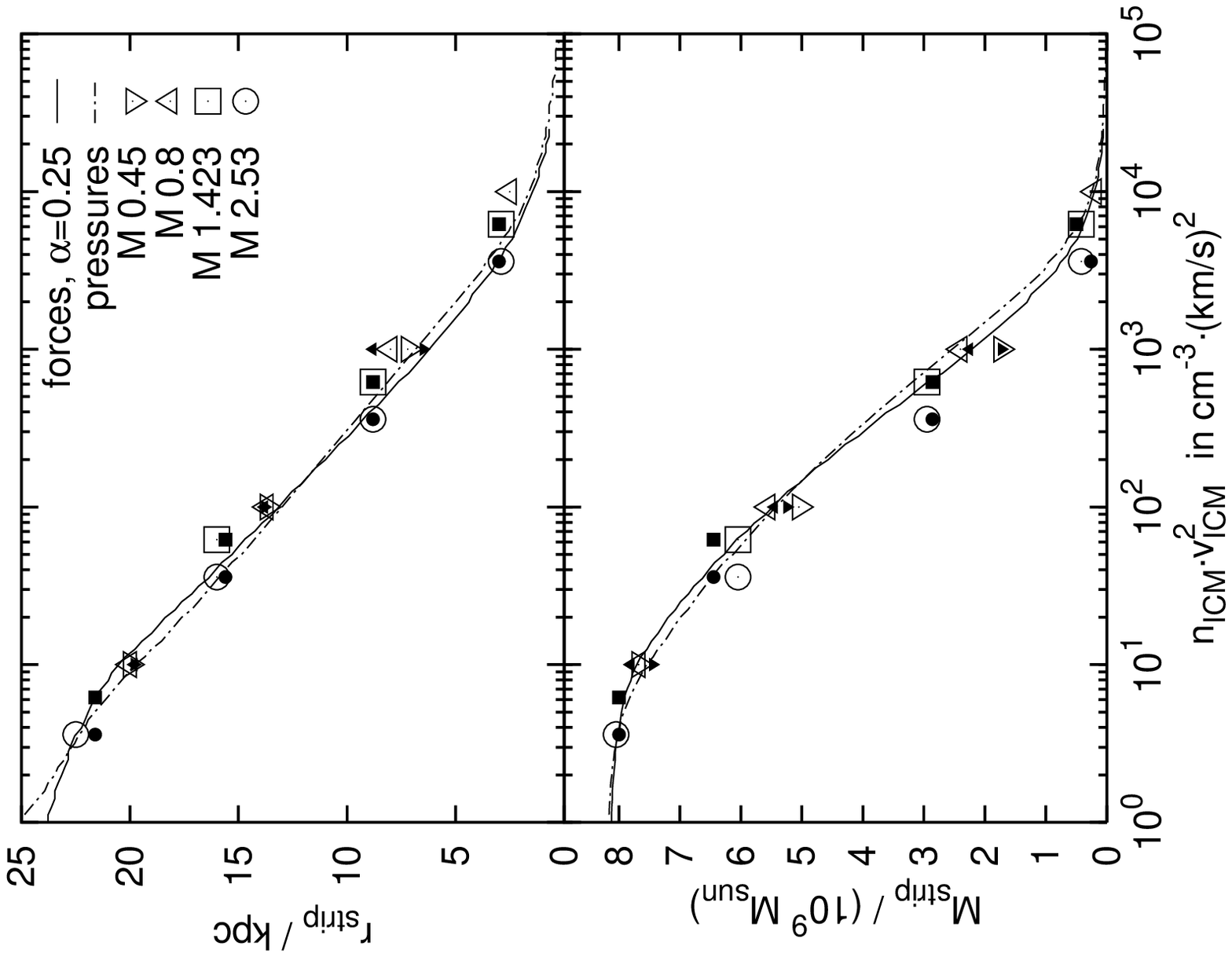}}
\caption{Comparison between the analytical estimates and the numerical 
result for the massive galaxy with Rankine-Hugoniot-corrected ram
pressures for the supersonic cases (see text).  Open and solid symbols
are for cases with $T\ICM{}_1$ and $T\ICM{}_2$, respectively. The
symbol shape codes the Mach number. The analytical estimate for the
force comparison (solid line) is fitted to the numerical results via
the parameter $\alpha$ (see also text). The estimate from the pressure
comparison is repeated unmodified.}
\label{fig:comp_an_num2}%
\end{figure}

\begin{figure}
\centering\resizebox{0.7\hsize}{!}{\includegraphics[angle=-90]{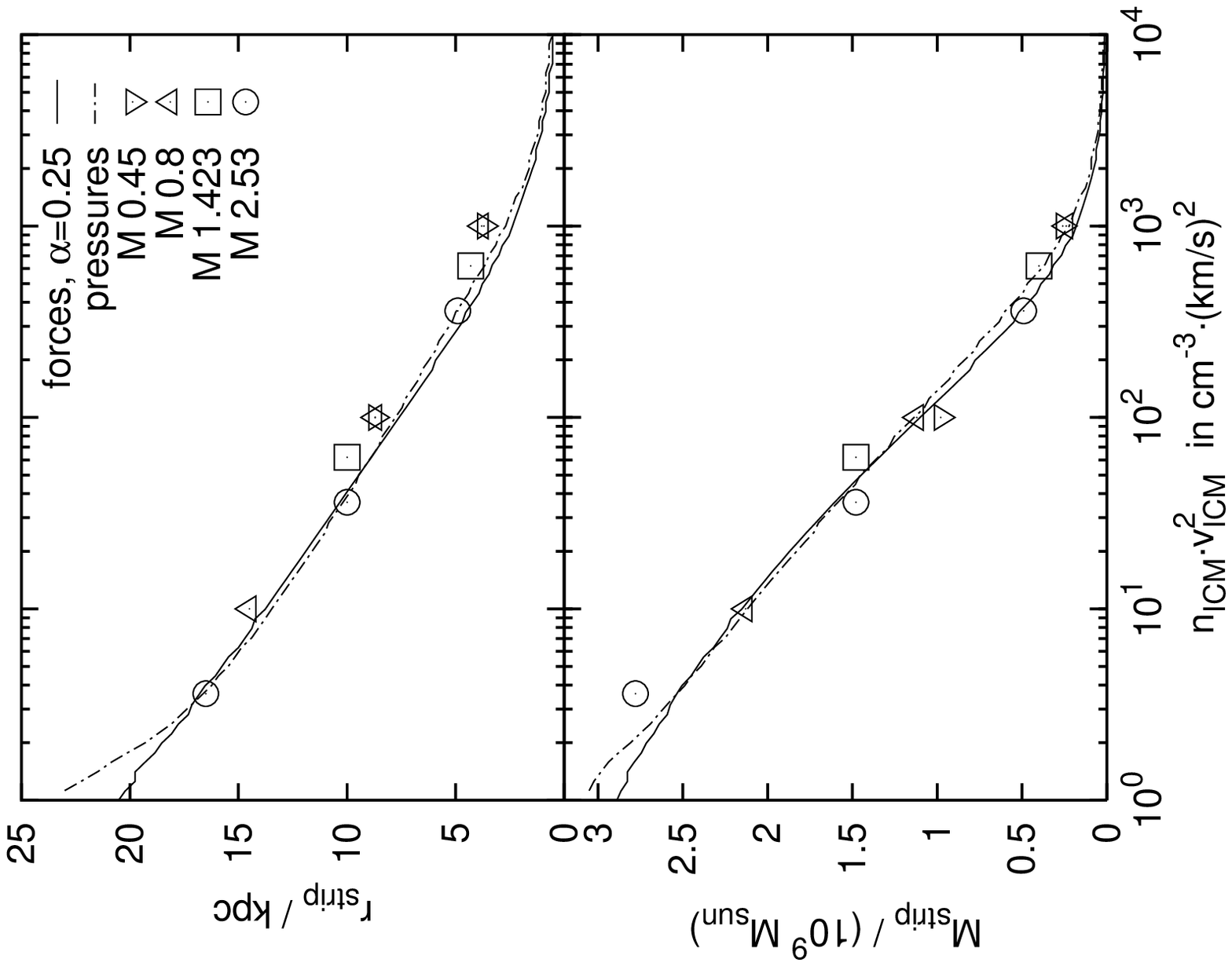}}
\caption{Same as Fig.~\ref{fig:comp_an_num2}, but for the medium galaxy.}
\label{fig:comp_an_num_medgal2}%
\end{figure}

We also want to consider the Mach number dependence of the numerical
results again. So far we calculated the ram pressure according to
$\tilde p\Ram=n\ICM v\ICM^2$ for the supersonic cases as
well. However, the velocity and density behind the formed bow shock
differ from the values on its upstream side. It may be more
appropriate to calculate the ram pressure for supersonic winds with
the values of $\rho$ and $v$ at the downstream side of a straight
shock with the corresponding Mach number. The component of the
momentum density $\rho\ICM v\ICM$ perpendicular to the shock front is
conserved across such a shock, whereas the velocity at the downstream
side is reduced (and the density increased) by a certain factor
according to the Rankine-Hugoniot conditions. Then also the true ram
pressure would be diminished by the same factor as $v$. For Mach
numbers of 1.423 and 2.53 (at the upstream side) the velocity at the
downstream side is reduced by factors of 0.62 and 0.36,
respectively. In Figs.~\ref{fig:comp_an_num2} and
\ref{fig:comp_an_num_medgal2} we replot the numerical results with
this correction, that is shifting the points for the supersonic
results down to the corrected ram pressures.
%
In addition, we repeat the analytical estimate from the pressure
comparison (version (d), marked with ``pressures'' in
Figs.~\ref{fig:comp_an_num2} and \ref{fig:comp_an_num_medgal2}) and
plot the estimate from the force comparison with an adjusted $\alpha$
(see Eq.~\ref{eq:fgrav3}, marked with ``forces, $\alpha=0.25$'' in
Figs.~\ref{fig:comp_an_num2} and \ref{fig:comp_an_num_medgal2}). For
$\alpha=0.25$ the numerical result and the analytical estimate agree
well. In contrast to the uncertainty expressed by $\alpha$ in the
force comparison, there is no uncertainty in the pressure
comparison. And indeed the (Rankine-Hugoniot-corrected) numerical
result and the estimate from the pressure comparison are in good
agreement. With $\alpha=0.25$, the estimate from the force comparison
was shifted to match the result from the pressure comparison. The
point for the lowest ram pressure and Mach 2.53 in
Fig.~\ref{fig:comp_an_num_medgal2} that lies above the analytical
estimate is explained easily because the dynamic intermediate phase
for this run is not finished at the end of the simulation. We note
that the correction of the ram pressures for the supersonic cases
according to the Rankine-Hugoniot conditions cannot describe the
situation completely, because it would mean that also in the
supersonic range the results should depend slightly on the Mach
number. This is not found in the simulations (see
Sect.~\ref{sec:result_comparison_TICM}).

\begin{figure}
\centering\resizebox{\hsize}{!}{\includegraphics[angle=-90]{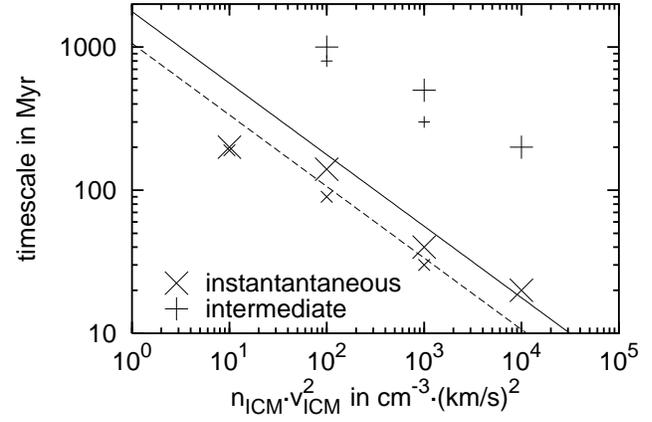}}
\caption{Duration of the instantaneous stripping phase and intermediate 
phase as a function of ram pressure. Large symbols are for the massive
galaxy, small symbols for medium galaxy. Also the analytical estimate
for the instantaneous phase according to Eq.~\ref{eq:t_strip} is
shown, the solid line is for the massive galaxy, the dashed one for
the medium galaxy.}
\label{fig:timescales}%
\end{figure}

We can also compare the predicted duration for the instantaneous
stripping with the numerical result. The prediction drawn from
Eq.~\ref{eq:t_strip} refers only to the replacement of the outer gas
disk. The numerical result for the replacement is given by the time of
the first local minimum in the $r\Disk$ curves in
Fig.~\ref{fig:comp-masses-radii}. The trends in the numerical
result and the analytical estimate shown in Fig.~\ref{fig:timescales}
agree well, also the absolute values agree roughly.
We also show the duration of the intermediate phase, that is the time
until all gas that is not bound tightly enough is truly unbound. We
find that this time is approximately 10 times as long as the
instantaneous stripping phase.

%
\subsubsection{Continuous stripping} \label{sec:result_comp_an-num_KH}
\begin{figure}
\resizebox{\hsize}{!}{\includegraphics[angle=-90]{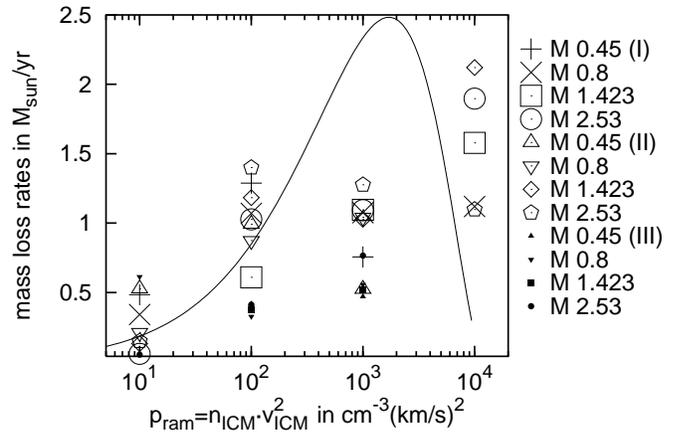}}
\caption{Mass loss rates of the continuous stripping phase as a function 
of ram pressure. The Mach numbers are coded by different
symbols. Symbols of group (I) (large symbols) are for the massive
galaxy in winds with $T\ICM{}_1$, group (II) (medium-size symbols) for
the massive galaxy in winds with $T\ICM{}_2$, and group (III) (small
symbols) for the medium galaxy in winds with $T\ICM{}_1$. We also show
the rough analytical estimate (see Eq.~\ref{eq:nulsen_new} in
Sect.~\ref{sec:analytic_continuous}).}
\label{fig:masslossrates_num}%
\end{figure}

The estimate of the mass loss rate during the continuous stripping
phase (see Sect.~\ref{sec:analytic_continuous}) is only a rough
one. The mass-loss rates are expected to be of the order of one
$M\Sun\,\Yr^{-1}$. Fig.~\ref{fig:masslossrates_num} demonstrates the
mass-loss rates derived from the simulations by fitting a linear
function to $M\Bound(t)$ during the continuous stripping phase.
Because the runs with $\tilde p\Ram=10\Rampresunit$ do not reach the
continuous phase during the runtime, for those runs we fit the time
range between 1 and $2\Gyr$. From the simulations we can derive
mass-loss rates between 0.1 and $2\,M\Sun\,\Yr^{-1}$. Except that the
highest ram pressures produced the highest mass-loss rates, we cannot
find any further trends. The analytical estimate could give a rough
impression only.

\section{Discussion and conclusions}
\label{sec:discussion}
%
\subsection{Influence of the ICM pressure}
What happens to the gas disk of a spiral galaxy in a high pressure
external medium? From the comparison of typical pressures in the disks
of isolated galaxies ($10^{-13}$ to $10^{-11}\Erg\,\ccm$) and typical
ICM pressures (e.g.~$1.4\cdot 10^{-13}\Erg\,\ccm$ for
$n\ICM=10^{-4}\ccm$ and $T\ICM=10^7\K$) alone it is obvious that
normal disk galaxies cannot exist in cluster environments, no matter
if they move or not. If for an isolated galaxy the external pressure
is increased gradually, the disk has two alternatives to react to
that. Either, if the density distribution should not change much, it
has to heat up in order to reach higher pressures so that it can
maintain pressure equilibrium with the surrounding gas. This is what
we assumed implicitly with our initial conditions. Or, if cooling
prohibits a substantial increase of the temperature, the disk will be
compressed. However, this is only the crude basic idea. The fate of a
more realistic multi-phase gas disk including effects like star
formation and feedback may be much more complicated. Due to the
compression the formation of molecular clouds and stars may be
enhanced, which in turn could increase the amount of heating by an
increased supernova rate and stellar winds. The observational fact
that even HI-deficient cluster galaxies contain a rather normal amount
of molecular gas \citep{kenney89,boselli97,casoli98} and show normal
to enhanced star formation rates in their inner part
\citep{koopmann98,koopmann04b,koopmann04a} supports this picture. It
would be worth to study the effect of the external pressure on the gas
disks of spiral galaxies in a separate work.

In how far does our setup of the initial model (fixing the density
distribution of the gas disk and calculating the temperature
distribution from hydrostatic equilibrium), which leads to high disk
temperatures, bias the results derived in this paper? The concern is
that due to the high temperature (and hence higher thermal energy) our
gas disks are not bound as strongly as cooler disks in isolated
galaxies and hence are stripped easier. In this case we would
overestimate the stripping efficiency. The alternative way of initial
setup would be to fix the temperature inside the gas disk
(e.g.~determined by the heating-cooling balance) as well as its mass
and calculate the density distribution from the hydrostatic
equation. The resulting gas disks would become thinner and have higher
densities as the external ICM pressure increases. In varying the gas
disk thickness and shape (see Sect.~\ref{sec:result_thickness}) we
have already checked if such thin disks are harder to strip by ram
pressure, but we did not find them to be more resistant to RPS. We
also performed test runs that started with a galaxy resembling an
isolated one (concerning disk pressure and temperature), but put into
a high-pressure ICM. As this galaxy was not in hydrostatic equilibrium
with the ICM, it was compressed immediately and then stripped like its
equilibrium analogue. Hence we conclude that our results are not
biased strongly by the initial pressure and temperature distribution.

\subsection{Morphological features of stripped galaxies}

In agreement with previous work, our simulations suggest that disk
galaxies suffering (approximately) face-on RPS should have truncated
disks, where the radius of the disk is set by the ram pressure and the
galactic potential. Moreover, if they are in the instantaneous
stripping phase or the dynamic intermediate phase, stripped gas could
be found at distances as far as $20\Kpc$ behind the disk. How long
such structures exist depends at least on $\tilde p\Ram$.  In how far
this stripped gas actually remains visible, is dispersed or cools,
forms stars or is affected by heat conduction, our simulations cannot
predict as they do not include such processes. Due to the homogeneous
gas disk used in our simulations, filaments emerging from the disk are
found mainly at the disk edge. Such examples are indeed observed
(e.g.~NGC 4522: \citet{vollmer04a} and references therein, NGC 4402:
\citet{cayatte90,cayatte94,crowl04}). The filaments found in NGC 4569
\citep{tschoke01,vollmer04,kenney04a,bomans04} seem not to emerge from
the edge but more from the centre of the disk. This may be due to
activity in the galactic nucleus or due to RPS of an inhomogeneous gas
disk.

\subsection{Mixing with the ICM}

We tried to infer to what extend ICM gets mixed into the remaining gas disk. 
Such mixing would change the angular momentum and metalicity of the
gas disk.  To investigate this point we used two more recipes to
compute the mass of the gas disk or the bound mass. In contrast to
summing over the ``coloured'' gas (the gas that has been inside the
galactic disk initially) in the disk region, we also summed up all gas
in the disk region with a temperature below $5\cdot 10^6\K$ (about the
virial temperature of the massive galaxy). This prevents that we
include the ICM that is streaming freely through the already swept
disk region, but it would include all ICM that has already mixed with
the gas disk. On the other hand, we sum over all gas in the region of
the bound gas instead of summing only over the ``coloured'' gas
there. This includes all ICM that is situated in this regions. We
find, however, no differences between the versions that should include
or exclude the ICM, so we conclude that no ICM is accreted by the
galaxy.

\subsection{Time dependent winds}

During its passage through a cluster a galaxy does not experience a
constant wind. Galaxies on more radial orbits fly through the high
density cluster core. With a typical velocity of $1000\Kms$ a galaxy
needs $1\Gyr$ to cross the inner Mpc of the
cluster. \citet{vollmer01a} calculated the time dependent ram pressure
for a galaxy falling into the Virgo cluster (their Fig. 3) and found
that typically the ram pressure exceeds $1000\Rampresunit$ for about
$100\Myr$ before and after the core passage, and still exceeds
$300\Rampresunit$ about $250\Myr$ before and after the core
passage. According to our simulations these times are definitely long
enough to truncate the radius of the gas disks. However, the duration
of the high ram pressure may be too short to unbind all this gas. Even
in our simulations with constant winds several $10^8M\Sun$ fall back
to the disk region, and for quite a while about $10^9 M\Sun$ of gas
that is still bound lingers behind the galaxy. If the ram pressure
abates too early after the core passage all this gas should fall back
to the disk. From our simulations we cannot infer in how far and how
fast the gas disk resettles after the core passage.

\subsection{Galaxies in cluster outskirts and galaxy groups}

In our simulations we found that even mild ram pressures of about
$100\Rampresunit$ and less already truncate the gas disk of massive
galaxies to about 15kpc. Even if low ram pressures do not unbind much
gas, they still displace the outer parts of the gas disk. The wind
itself and back-falling gas disturb the disk, send shock waves through
it which could trigger star formation \citep[e.g.~][]{fujita99}. We
therefore expect that even at larger distances from the cluster centre
the gas disks of most galaxies should be ``pre-processed'' \citep[see
also][]{fujita04}. This pre-processing of galaxies in cluster
outskirts and groups could explain the HI deficient galaxies at large
distances to the cluster centre found by \citet{solanes01}. RPS in low
density environments may be more important than thought so far.

\subsection{Inclination}
We have studied only cases where galaxies move face-on. What can we
learn for inclined cases? \cite{marcolini03} found that the
inclination angle matters only for medium ram pressures. This can be
understood also in the context of our simulations. For very weak winds
hardly any gas is unbound from the galaxy by the pushing effect, the
main mechanism is the continuous stripping. For this mechanism the
inclination angle is not important, it works on the whole surface. For
very strong winds the complete disk is lost independently of
inclination. Only for the intermediate winds the inclination decides
how much mass can be pushed out in the beginning, hence for heavily
inclined cases less gas is lost in this phase.  After that the
continuous stripping continues at the stage where the previous phase
has ended. Also \citet{quilis00} agreed that only strict edge-on
galaxies can retain more gas, intermediately inclined galaxies are
stripped similarly to face-on ones. This is also in agreement with the
results of \citet{vollmer01a}, who constructed a generalised formula
relating the remaining gas mass with the ram pressure and the
inclination angle. However, their sticky-particle simulations cannot
model the turbulent viscous stripping. \citet{schulz01} found that for
inclined galaxies less gas lingers in the lee of the disk, hence our
results concerning back-falling gas may change in such
situations. However, the unbinding of the gas after it has left the
disk region should take some time also for inclined cases.

In addition to the restriction to the face-on case, a 2D code is
restricted to rotationally symmetrical cases. However, real systems
are not symmetrical, and an asymmetry may change the behaviour of
turbulence and instabilities.

\subsection{Summary}  \label{sec:summary}
We performed a large set of numerical simulations of the face-on RPS
of disk galaxies in constant ICM winds. We find that even massive
galaxies can lose a substantial part of their gas disks, and can be
stripped completely in cluster centres. Our simulations also covered
low density environments like cluster outskirts, and we find that
there galaxies are truncated to 15 to $20\Kpc$. Back-falling gas and
the compression from the flow could trigger star formation, hence we
expect galaxies to be pre-processed in cluster outskirts and groups as
well.  Furthermore, the models show that the stripping proceeds in
three phases:
\begin{itemize}
\item Instantaneous stripping phase: the outer disk is pushed out of its 
      initial position (bent towards the downstream side), but it is
      only partly unbound. The main radius reduction happens in this
      phase. This phase is short (20 to $200\Myr$), its duration as a
      function of ram pressure can be read from
      Fig.~\ref{fig:timescales}.
\item Intermediate phase: rather dynamic, the gas outside the disk region 
      partially falls back to the disk (about 10\% of the original gas
      mass), the rest is completely stripped. This may repeat in
      cycles. This phase lasts about 10 times as long as the
      instantaneous stripping phase, it is finished when no
      substantial amount of bound gas is left outside the disk
      region. The radius does not change much during this phase, but
      the bound gas mass and also the gas mass inside the disk region
      do. The mass and radius of the remaining gas disk at the end of
      this phase can be predicted best by comparing the thermal
      pressure in the disk plane and the ram pressure (for details see
      Sect.~\ref{sec:analytic_instantaneous},
      Eq.~\ref{eq:estimate_comp_pres}). The analytical estimate and
      the numerical results for the stripping radius and the remaining
      gas mass are summarised in Figs.~\ref{fig:comp_an_num2} and
      \ref{fig:comp_an_num_medgal2} for the massive and the medium
      mass galaxy, respectively.
\item Quasi-stable continuous stripping: constant small mass loss rate 
      of about $1M\Sun\,\Yr^{-1}$, slow decrease of the disk radius.
\end{itemize}
The main parameter that sets the mass and radius of the remaining gas
disk is the ram pressure, but there is a slight dependence on Mach
number in the sense that in the subsonic regime and at the transition
from the subsonic to the supersonic regime ICM flows of lower Mach
numbers result in a bit stronger gas loss than flows with higher Mach
numbers. In how far this result holds for a two-fluid treatment of the
ICM is unclear.

The mass loss and shrinking of the radius depend only on the surface
density of the gas disk, but not on its vertical structure (thickness,
flared or exponential).

We observe a compression at the upstream side of the remaining disk,
but we do not find an increased surface density.

\begin{acknowledgements}
This work was supported by the \emph{Deut\-sche
For\-schungs\-ge\-mein\-schaft\/} project number He~1487/30. We
gratefully acknowledge fruitful and helpful discussions with Christian
Theis, Joachim K\"oppen, Bernd Vollmer and Curtis Struck. We also
thank the anonymous referee for the helpful comments.
\end{acknowledgements}

\appendix
\section{Influence of artificial viscosity}
\label{sec:viscosity}
\begin{figure}
\resizebox{\hsize}{!}{
\includegraphics[height=\textwidth,angle=-90]{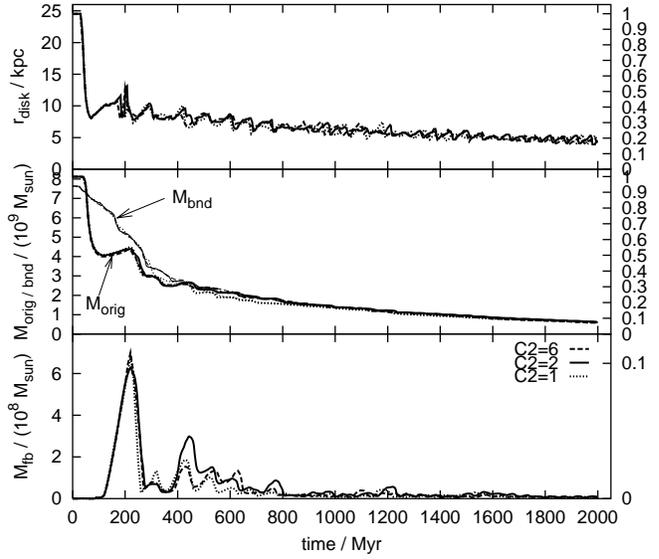}}
\caption[Comparison of gas disk radius, mass and fallen-back mass for different viscosities]%
{Influence of the viscosity on the evolution of the gas disk radius, mass and
fallen-back mass. The value of the viscosity parameter $C_2$ (see
Sect.~\ref{sec:numerics}) is indicated in the
key. For further parameters of these runs see Table~\ref{tab:simulations_visc}. This plot is for the runs with Mach number 0.8.}
\label{fig:compare-evol-visc1}
\end{figure}

\begin{figure}[!tb]
\resizebox{\hsize}{!}{
\includegraphics[width=\textwidth,angle=0]{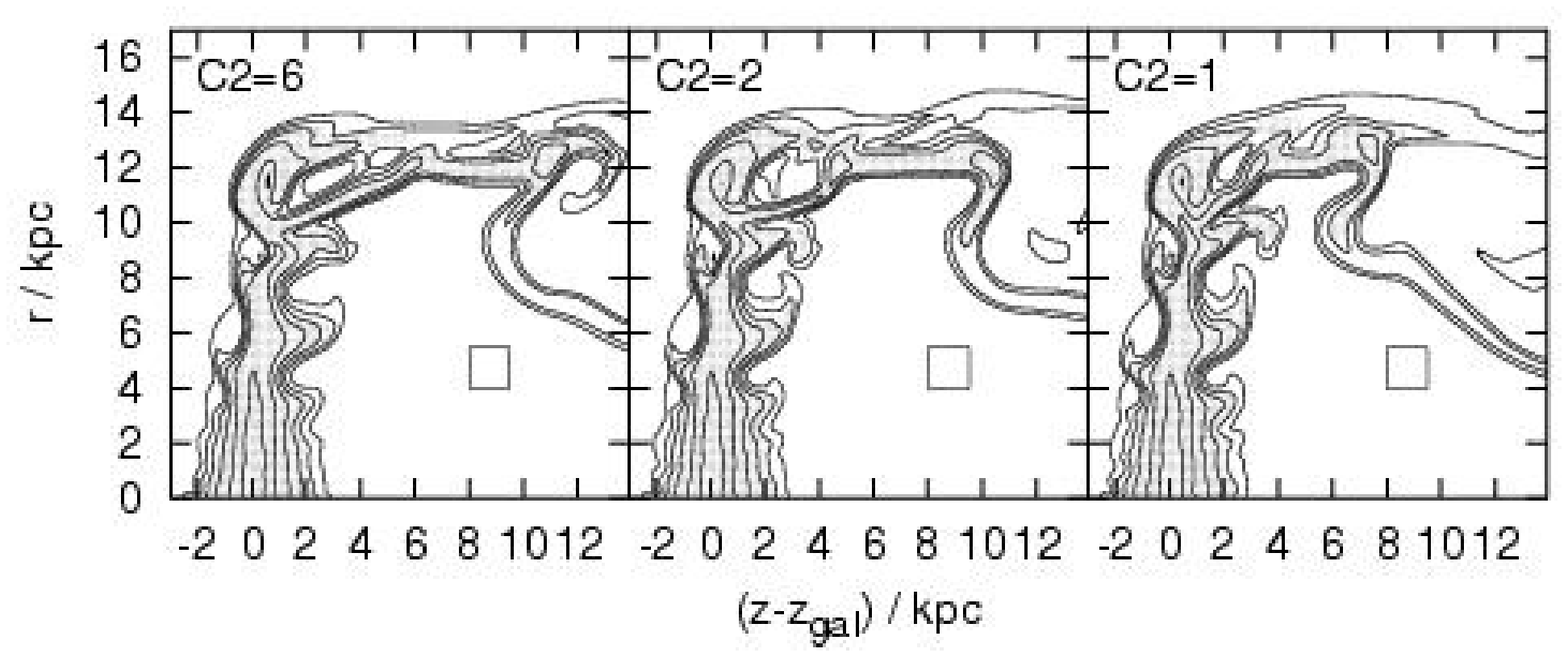}}
\resizebox{\hsize}{!}{
\includegraphics[width=\textwidth,angle=0]{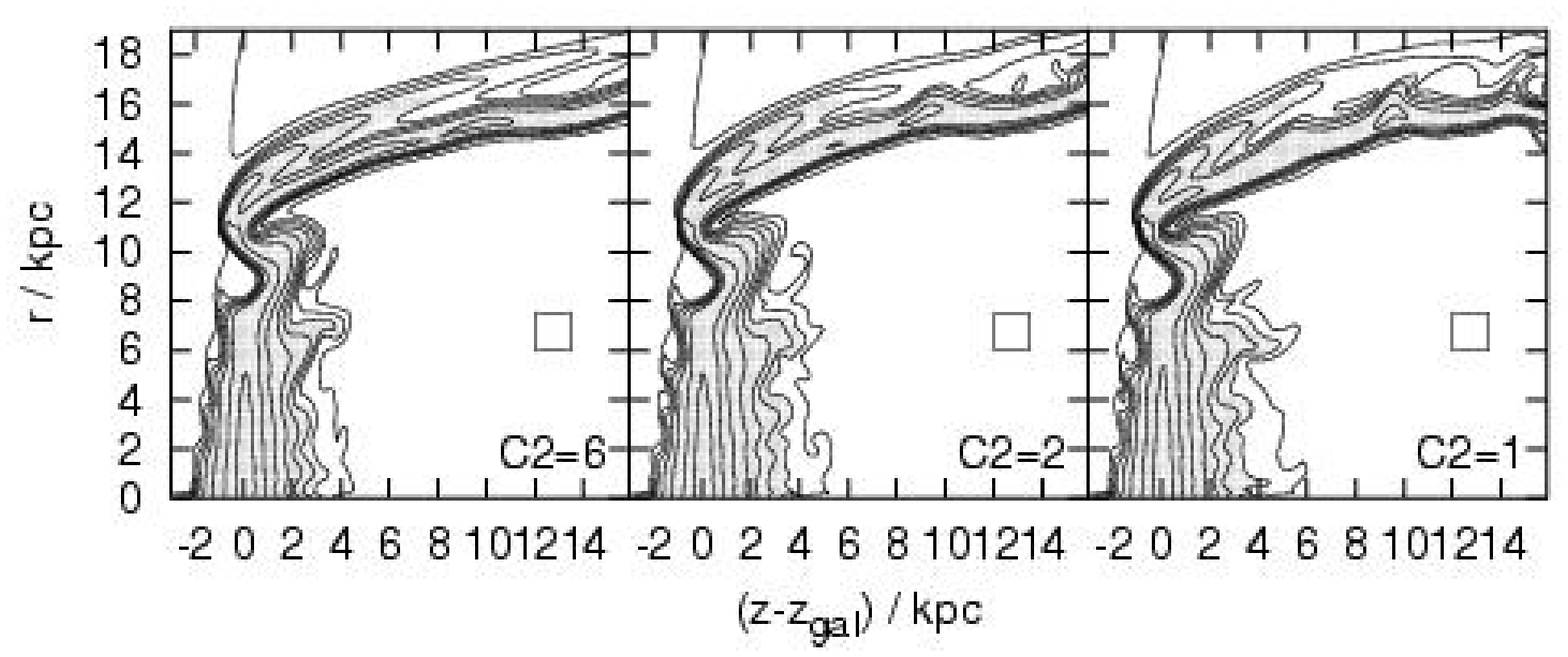}}
\caption[Comparison of the density distribution for different viscosities]%
{Comparison of the density distribution (at $t=200\Myr$) for different
viscosities. The plot style is the same as in
Fig.~\ref{fig:contours}. The top panels are for Mach number 0.8, the
bottom panels for Mach number 2.53. For further parameters see
Table~\ref{tab:simulations_visc}.  Each panel is labelled with the
appropriate viscosity parameter $C_2$ (see
Sect.~\ref{sec:numerics}). The rectangle corresponds to $10\times 10$
cells.}
\label{fig:compare-contours-visc2}
\end{figure}

In our simulations we observed the long phase of turbulent viscous
stripping. The code uses an artificial viscosity mainly to be able to
handle shocks. A concern may be that due to this artificial viscosity
the code cannot handle the KH-instability correctly and hence biases
the results of the viscous stripping phase. We performed test runs
(see Table~\ref{tab:simulations_visc}) to investigate the influence of
the artificial viscosity by varying the viscosity parameter $C_2$ (see
Sect.~\ref{sec:numerics}).
\begin{table}[!tb]
\centering\begin{tabular}{ccc}
\hline
$\tilde p\Ram$ &  Mach number & viscosity parameter                     \\
($\Rampresunit$)&              & $C_2$ (see Sect.~\ref{sec:numerics}) \\
\hline\hline
1000           &  0.8         & 1,2,6 \\
$\cdot$        &  2.53        & $\cdot$ \\
\hline
\end{tabular}
\caption{List of simulations to study the influence of the artificial
viscosity. Common parameters for all runs are: massive galaxy with
standard exponential disk, ICM wind with $T\ICM{}_1$
($c\ICM=1000\Kms$).}
\label{tab:simulations_visc}
\end{table}
For the results for the disk radius, mass and bound mass the differences between runs with various viscosities are negligible. We show one case in Fig.~\ref{fig:compare-evol-visc1}.
%
%
Fig.~\ref{fig:compare-contours-visc2}
compares snapshots at $t=200\Myr$ for the test cases with varying
viscosity. 
%
%
As expected, for the high viscosity parameter, the surface
instabilities are softer, whereas for in the low viscosity case
($C_2=1$) the features are sharper. However, almost all features can
be identified for all viscosities. We conclude that the use of the
artificial viscosity does not bias our results.

\section{Influence of resolution}
\label{sec:resolution}
In order to test the influence of the resolution we have performed
test runs with resolutions $50\PC$, $100\PC$, $150\PC$ and $200\PC$. 

\begin{figure}[!tb]
\resizebox{\hsize}{!}{
\includegraphics[height=\textwidth,angle=-90]{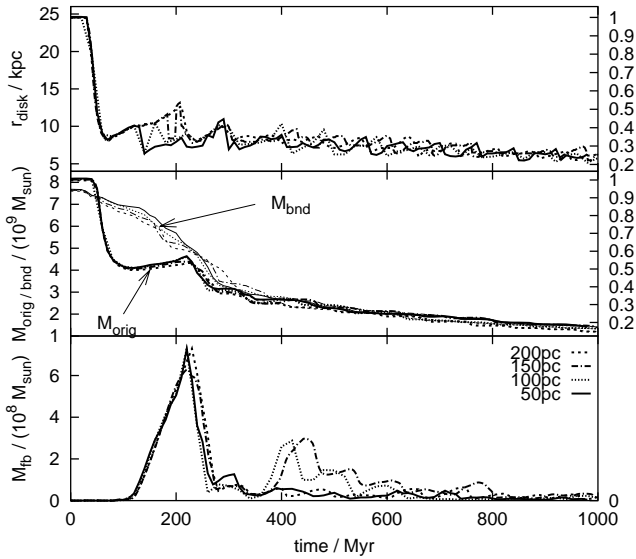}}
\caption{Influence of grid size and resolution on evolution of gas disk
radius, mass and fallen-back mass. See
text for parameters.}
\label{fig:compare-evol-gridsize-resolution}
\end{figure}

Fig.~\ref{fig:compare-evol-gridsize-resolution} compares the radius $r\Disk$
and mass $M\Disk$ of the remaining gas disk for the standard run (massive
galaxy, exponential gas disk, $b\Gas=0.4\Kpc$, $t\ICM{}_1$, $\tilde
p\Ram=1000\Rampresunit$, Mach number 0.8) for the four different
resolutions. 
%
%
Also the the mass of bound gas $M\Bound$ and the mass of
fallen-back gas $M\Fallback$ are shown. The differences for all
quantities are negligible. The small differences in $M\Bound$ arise
because in cases with higher resolution the the stripped gas can be
compressed into smaller volumes, reaching higher densities
locally. Such dense clouds are more likely to be bound than the clouds
with lower densities. 

\begin{figure}[!tb]
\resizebox{\hsize}{!}{\includegraphics[width=\textwidth,angle=0]{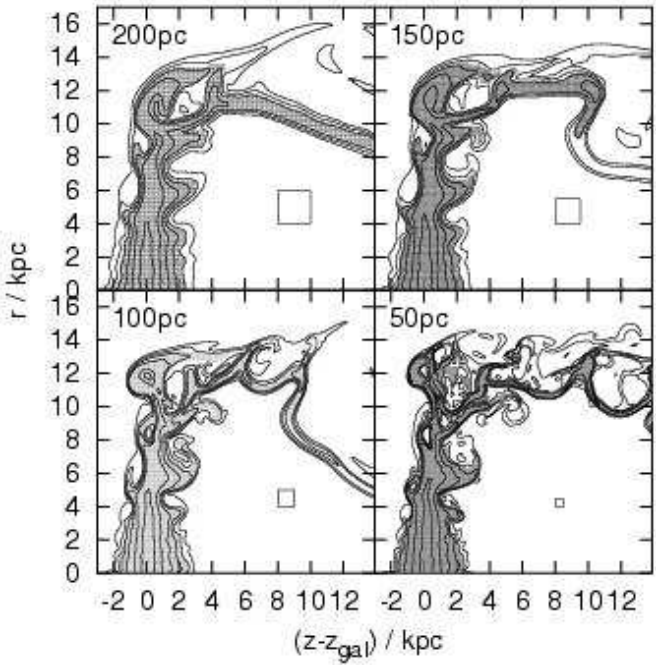}}
\caption[Comparison snapshots for different resolutions]%
{Influence of resolution: Snapshots at $t=200\Myr$ for four different
resolutions as given in the plot, see text for parameters. The plot
style is the same as in Fig.~\ref{fig:contours}. The size of $10\times
10$ cells is indicated by the rectangle in each panel.}
\label{fig:compare-snapshots-resolution}
\end{figure}

In Fig.~\ref{fig:compare-snapshots-resolution} we show snapshots at $t=200\Myr$ for the four different resolutions.
%
%
The strongest difference occurs for the stripped material. The coarser
grids cannot resolve the small-scale fragmentation of the stripped gas
evident in the highest resolution. However, we did not attempt to
interprete the fate of the stripped material shown in the simulations,
as we neglected e.g.~thermal conduction, which might be play an
important role in the evolution of the stripped gas.

%
\bibliographystyle{aa}
\bibliography{%
/home/elke/ARBEIT/TEXT/TEXTTOOLS/theory_simulations,%
/home/elke/ARBEIT/TEXT/TEXTTOOLS/hydro_processes,%
/home/elke/ARBEIT/TEXT/TEXTTOOLS/numerics,%
/home/elke/ARBEIT/TEXT/TEXTTOOLS/observations_general,%
/home/elke/ARBEIT/TEXT/TEXTTOOLS/observations_clusters,%
/home/elke/ARBEIT/TEXT/TEXTTOOLS/observations_galaxies,%
/home/elke/ARBEIT/TEXT/TEXTTOOLS/galaxy_model,%
/home/elke/ARBEIT/TEXT/TEXTTOOLS/gas_halo,%
/home/elke/ARBEIT/TEXT/TEXTTOOLS/icm_conditions,%
/home/elke/ARBEIT/TEXT/TEXTTOOLS/else}

\end{document}